\documentclass{aa}

\usepackage{aalongtable}
\usepackage{graphicx,amssymb,verbatim,natbib,rotate,lscape,supertabular}

\bibpunct{(}{)}{;}{a}{}{,}

\usepackage{graphicx} \usepackage{txfonts}
\usepackage[latin1]{inputenc} \usepackage{color}
\usepackage{natbib}

\begin{document} 

   \authorrunning{Poncelet et al.}

   \title{An original interferometric study of NGC~1068 \\ with VISIR
   BURST mode images\thanks{Based on commissioning time observations collected at the ESO/Paranal MELIPAL
   telescope.}}

   \institute{LUTH, Observatoire de Paris, 92195 Meudon Cedex\ \and
     LESIA, Observatoire de Paris, 92195 Meudon Cedex\ \and
     CEA/DSM/DAPNIA/Service d'Astrophysique, CE Saclay F-91191
     Gif-sur-Yvette \\ \email{anne.poncelet@obspm.fr} }

   \author{A.  Poncelet \inst{1,2}, \ C. Doucet\inst{3}, \
   G. Perrin\inst{2}, H. Sol\inst{1} \and  P.O. Lagage\inst{3} }

   \date{Received , 2006; accepted , 2007}

\abstract{We present 12.8~$\mu$m images of the core of NGC~1068, the
  archetype Seyfert type II galaxy, obtained during first operations
  of the \textit{BURST mode} of the VLT/VISIR (Imager and Spectrometer
  in the InfraRed at the Very Large Telescope).}
{We trace structures under the diffraction limit of one UT (Unit
  Telescope at the VLT) and we investigate the link between dust in
  the vicinity of the central engine of NGC~1068, recently resolved by
  interferometry with MIDI (Mid-InfrareD Interferometer), and more
  extended structures. This step is mandatory for a multi-scale
  understanding of the sources of mid-infrared emission in active
  galactic nuclei (AGN).}
{A speckle processing of VISIR \textit{BURST mode} images was performed to
  extract very low spatial-frequency visibilities, first considering
  the full field of VISIR \textit{BURST mode} images and then limiting it to
  the mask used for the acquisition of MIDI data.}
{Extracted visibilities are reproduced with a multi-component
  model. We identify two major sources of emission at 12.8~$\mu$m: a
  compact one $<$~85~mas, directly associated with the \textit{dusty
  torus}, and an elliptical one of size ($<$~140)~mas~$\times$~1187~mas and P.A.~$\sim$~-4$\degr$ (from
  north to east), which gives a new description of the NS elongation
  of the nucleus. This is consistent with previous deconvolution processes.

The combination with MIDI data reveals the close environment of the
  \textit{dusty torus} to contribute to $\sim$~83$\%$ of the MIR flux seen by MIDI.}
{This strong contribution has to be considered in modeling
 long baseline interferometric data. It must be related to the
 NS elongated component which is thought to originate from individually
 unresolved dusty clouds and is located inside the ionization cone
 where it is photoevaporating and radiatively accelerated. Low temperatures of the
 \textit{dusty torus} are not challenged, emphasizing the scenarios of
 clumpy torus.}

\keywords{galaxies: NGC 1068 -- galaxies: Seyfert -- galaxies: active
  galactic nuclei -- infrared: galaxies -- techniques: imager -- instrument: VISIR --
  techniques: speckle interferometry}

\maketitle

\section{Introduction} 

NGC~1068, a bright and near archetype Seyfert type II galaxy is a key
target in testing the unified scheme of active galactic nuclei (AGN). This stipulates that
Seyfert II and Seyfert I are the same objects, depending upon the inclination
of the nucleus along the line of sight \citep{Antonucci_1985}. The
\textit{dusty torus} is therefore an important element of this
paradigm. The most recent theoretical models are given by
\citet{Nenkova_2002}, \citet{Schartmann_2005}, \citet{Honig_2006},
\citet{Elitzur_2006}, among others.

Multi-wavelength observations of the nucleus of NGC~1068 have been reviewed by
\citet{Galliano_2003}.  Recently, deconvolution of near-IR adaptive
optics (AO) images of NGC~1068 shows an unresolved core of some 30~mas
along the NS axis and $<$~15~mas along the EW axis (in the K$_s$
band). Additionally, upper size limits of 50~mas in the L$'$ and M$'$
bands were set \citep{Gratadour_2006}. These data also exhibit an
$\sim$~800~mas extended, peculiar S-shaped distribution aligned with
the radio jet. Near-IR speckle interferometry has revealed a
component of FWHM~$\sim$~18~mas~$\times$~39~mas oriented at
PA~$\sim$~-16$\degr$ \citep{Weigelt_2004}. Combined with the VLT/VINCI measurement of
\citep{Wittkowski_2004}, data were fitted with two components of
$FWHM$~$<$~5~mas and 40~mas. The first MIR images reported by
\citet{Bock_2000} traced the same NS elongation of the nucleus, and the
12.8~$\mu$m deconvolved images from the VLT/VISIR
\citep{Galliano_2005} highlights several knots distributed along the
S-shape (twisting from PA~=~-4$\degr$ to PA~=~31$\degr$). Using
mid-IR spectroscopy, \citet{Mason_2006} distinguished two sources of
mid-IR emission, a compact core within the central 4$''\times$4$''$
and a more extended component linked to heated dust inside the
ionization cone. Mid-IR interferometric observations with the VLT/MIDI
were necessary to resolve the \textit{dusty torus} of NGC~1068. The first
data were obtained in 2003 and presented by \citet{Jaffe_2004}. Two
independent analyses have revealed a compact structure of dust composed of
amorphous silicates, with size $<$~82~mas and relatively low temperatures 
\citep[on the order of 350~K;][]{Jaffe_2004,Poncelet_2006}. 

Configurations of Unit Telescopes (UTs) at the VLTI do not provide baselines shorter than 30~m, although low spatial-frequency
visibility points would yield tight constraints for modeling MIDI data. One
way to at least partially solve this problem is observe with a single UT
to access baselines between 0 and 8.2~m (the pupil of one UT) using an
interferometric processing of images. 

NGC~1068 was observed at 12.8~$\mu$m in January 2005 using the
\textit{BURST mode} (or quasi-Speckle mode) of VISIR. The big advantage of
this is the shorter acquisition time of 16 milliseconds, far shorter
than the typical atmospheric turbulence variations of $\sim$~300~ms at
10~$\mu$m. The aim of the study is therefore to present
the \textit{BURST mode} of VISIR to show its advantages and its use for the
speckle analysis. This allows us to derive the mid-IR low spatial-frequency visibilities between 0 and 8~m of baseline for the first time. Next we compare
these to MIDI data points in order to establish the
link between the source of mid-IR emission on different scales.
Section~\ref{observations} gives details on the \textit{BURST mode} of VISIR
and presents the principle of how visibilities are extracted from
VISIR images. Results of the modeling applied to the VISIR
visibilities are shown and discussed in
Sect.~\ref{study}. Section~\ref{comparison} presents the link between
VISIR and MIDI observations. Section~\ref{conclusion} gives the
general conclusions drawn by this study.

\section{Observations and data processing} \label{observations}

\begin{figure*}
  \begin{center}
   \includegraphics[height=4.53cm]{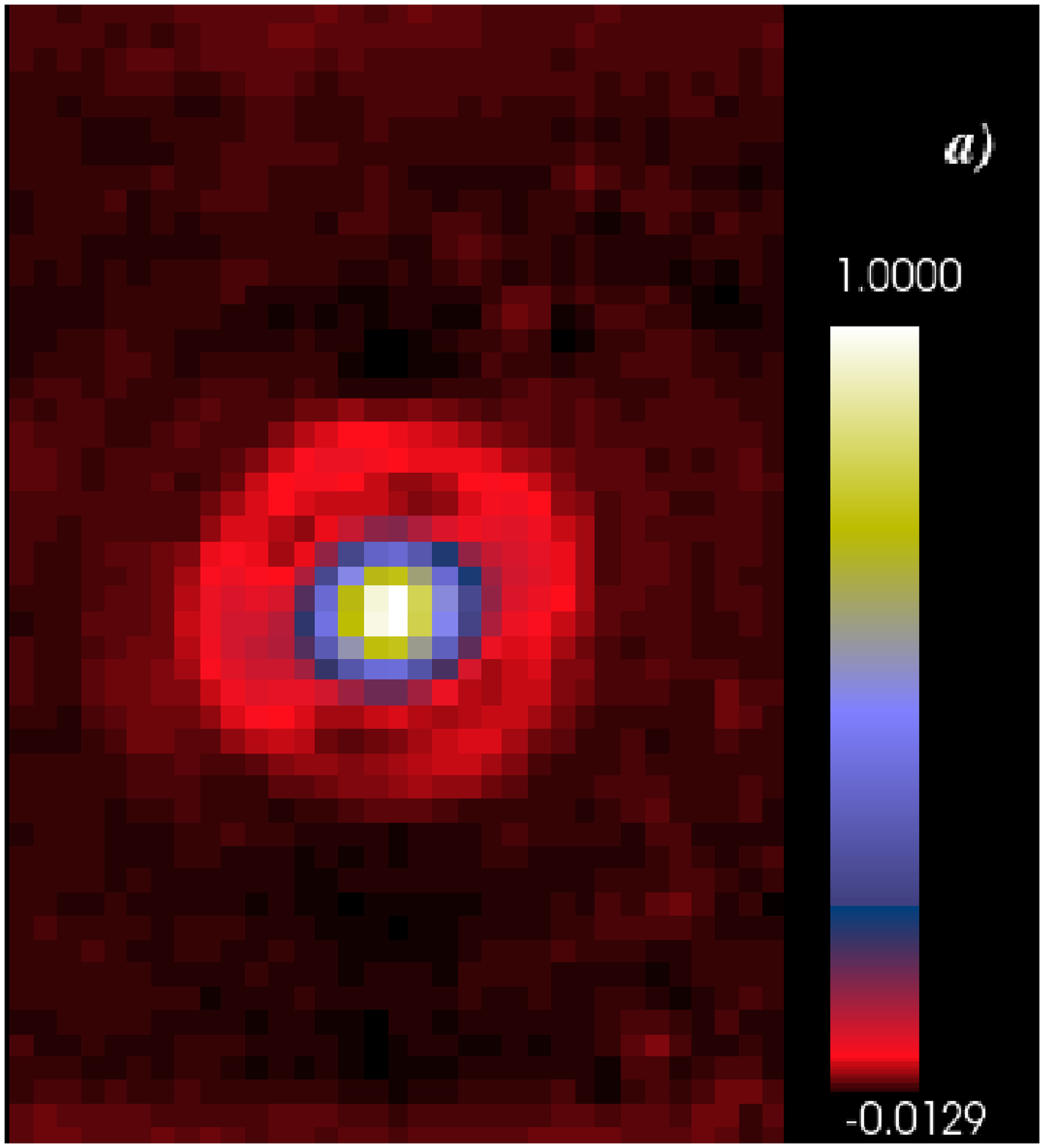}
   \includegraphics[height=4.53cm]{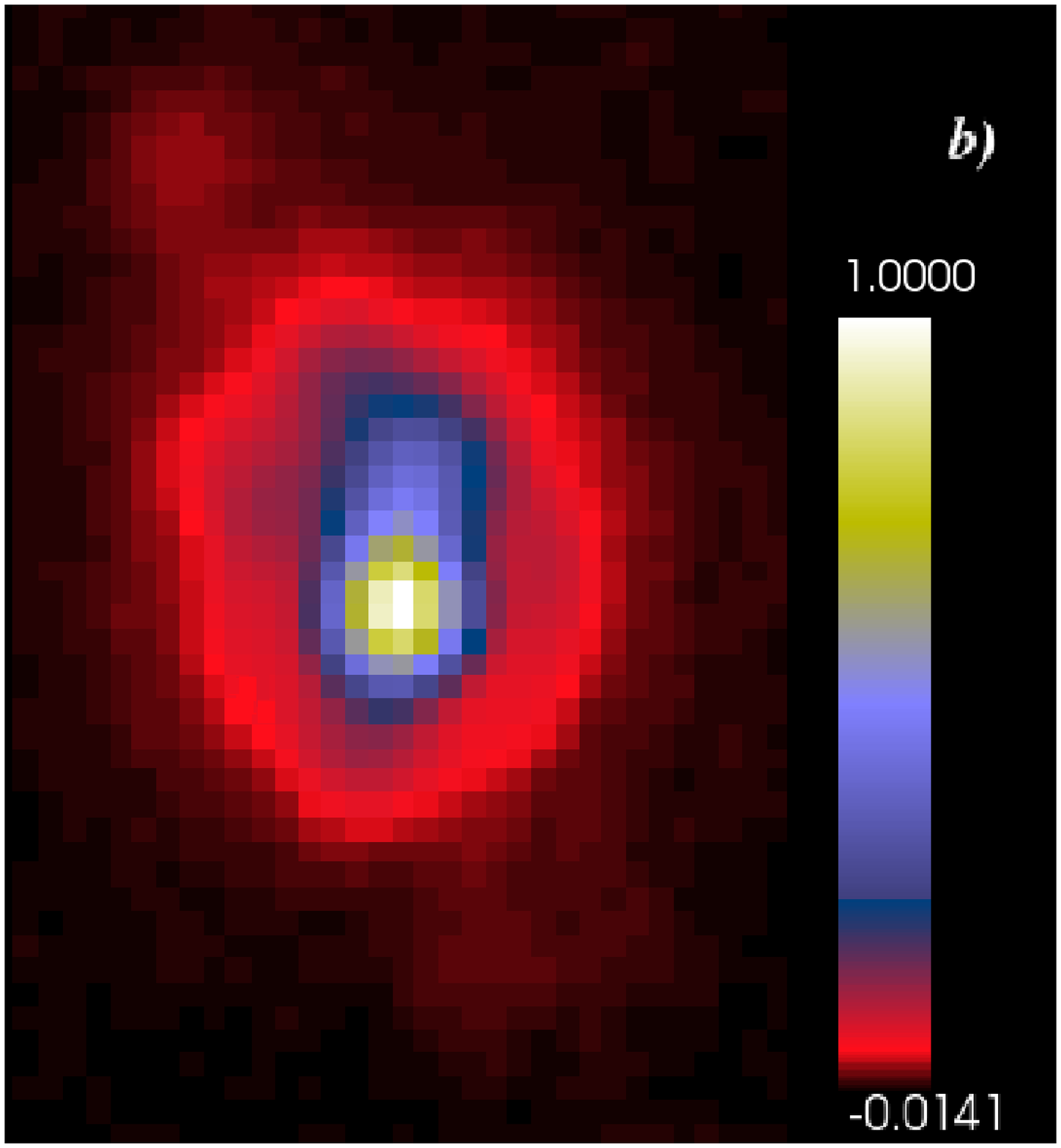}
   \includegraphics[height=4.5cm]{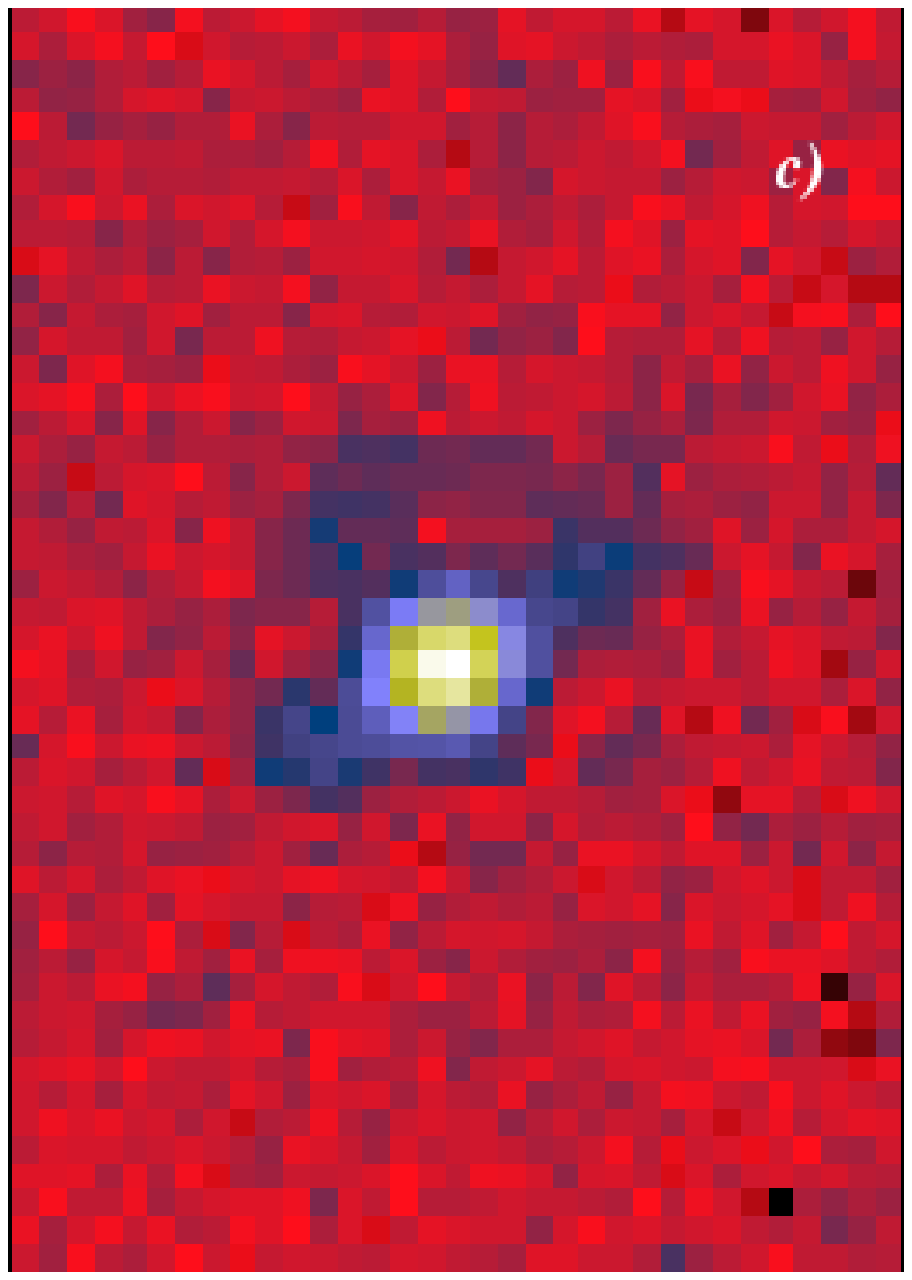}
   \includegraphics[height=4.5cm]{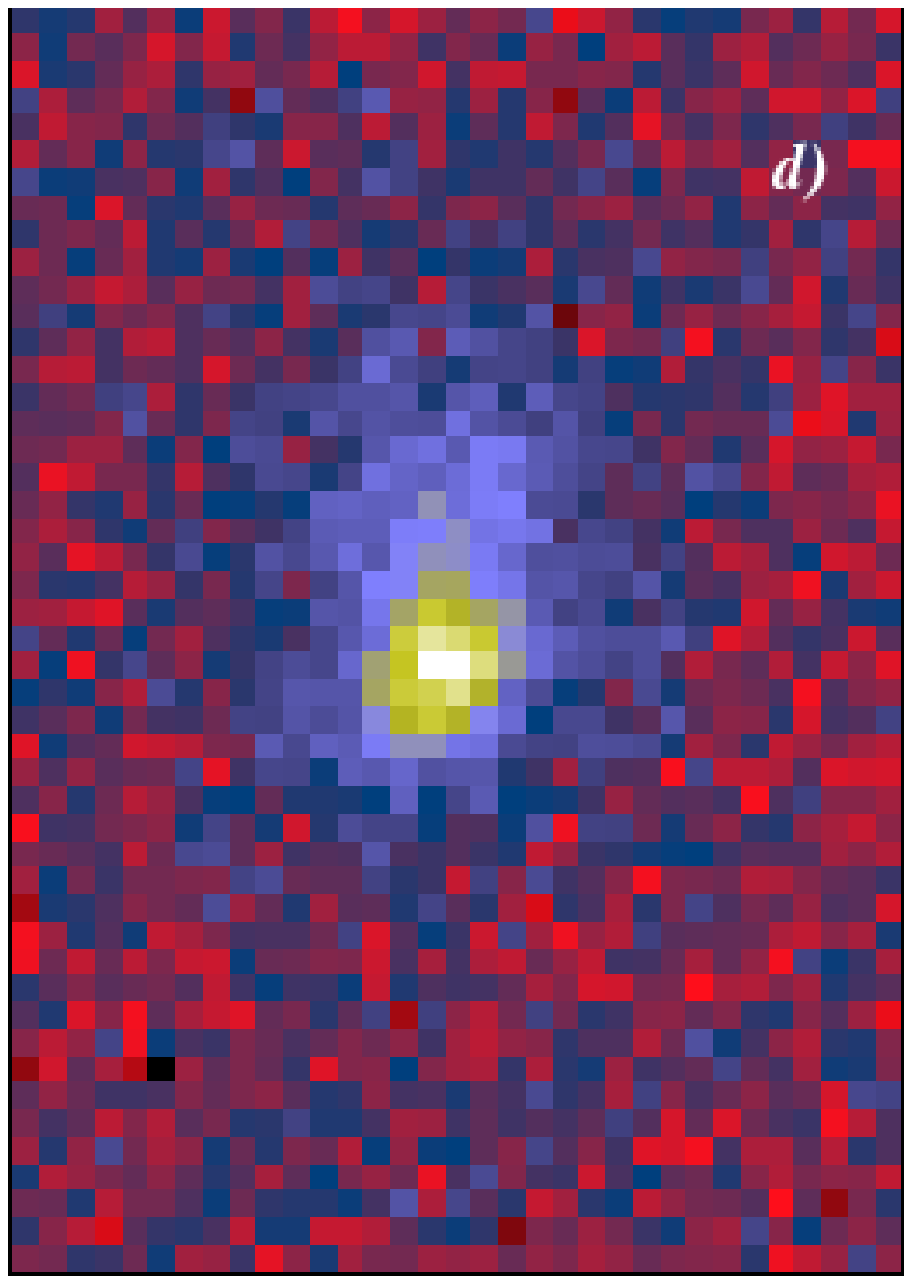}
  \end{center} 
 \caption{12.8~$\mu$m images obtained with the \textit{BURST mode} of
 VISIR. For comparison, normalized images are displayed with the
 same color table. \textit{From left to right}: $a)$ Sum
 of the 2200 individual exposures on HD~11383 (reference star). The good signal-to-noise ratio has
 to be compared to individual exposures. Here we clearly see the Airy
 disk ($\sim$~10.5 pixels large centered on the brightest pixel).
 $b)$ Sum of the 13000 exposures obtained on NGC~1068. The full
 structure appears clearly:
 the NS elongation from the unresolved compact core, the asymmetric
 emission and a knot farther away toward the NE (north is up and east
 is left). The appearant EW elongation is due to the diffraction ring
 at $\sim$~393~mas from the core. $c)$ Example of an individual 16~ms exposure of
 HD~11383. The Airy disk faintly appears, attesting to the good
 seing conditions during the night of observation. $d)$ An individual exposure of
 NGC~1068, already showing the NS elongation of the nucleus. 
 }
 \label{images}
\end{figure*}

NGC~1068 was observed on the night of January 25, 2005 using the
\textit{BURST mode} of VISIR, which is installed
on the VLT (Paranal, Chili). The actual observation mode used is
described in \citet{Doucet_2006}. Total exposure time on the object
was 208 seconds using the 12.8~$\mu$m Ne~II filter ($\Delta
\lambda$~=~0.21~$\mu$m). The exposure time was 16~ms for individual
images. The airmass was $\sim$~1.20 at the
time of observations. HD~11383, a K5III spectral type star, was
observed as a calibrator. To correct for the turbulence with offline processing, data are stored every 1000 elementary images
for one nodding position, using a chopping frequency of 0.25~Hz in the
NS direction. After classical data reduction in the mid-IR range, a
cube of 500 images, both chopped and nodded (4 beams/image), is
obtained. Turbulence has the effect of moving each of the 4 sources
independently so they have to be individually extracted from the
cube. Finally images of similar quarter are shifted and added,
producing some 13000 and 2200 elementary images of NGC~1068 and
HD~11383, respectively. To obtain a better signal-to-noise ratio -- and
according to the good correlation between elementary images -- they are
added in blocks of 5 (the sums may be considered as instantaneous maps
of the source intensity). Individual and resulting 12.8~$\mu$m VISIR
\textit{BURST mode} images of NGC 1068 and of HD~11383 may be found in Fig.~\ref{images}.

The \textit{BURST mode} is important because it uses such a short
exposure time that it effectively freezes atmospheric
turbulence. Therefore it enabled us to apply the speckle technique first introduced by
\citet{Labeyrie_1970} as an analytical tool. This aims to reach the diffraction limit of a large telescope (i.e. $\sim$~320~mas at
12.8~mas), after incoherent integration of exposures.

The seeing was excellent and very stable during observations of
NGC~1068 and images are diffraction-limited. Thus, the monochromatic
transfer function $\tilde{T}_\lambda(u,v)$ is estimated well by
observations of the calibrator star. Unbiased visibilities for
NGC~1068 and for the calibrator star are computed from the power
spectra of the images:
 
\begin{eqnarray}
 \vert V_\mathrm{~calib}(u,v) \vert ^2 & = & \langle \vert
 \tilde{I}_\mathrm{~calib}~(u,v) \vert ^2 \rangle_t - \langle \vert
 \tilde{N}_\mathrm{~calib}~(u,v) \vert ^2 \rangle_t \\
 \vert V_\mathrm{~NGC1068}~(u,v) \vert ^2 & = & \langle \vert
 \tilde{I}_\mathrm{~NGC1068}~(u,v) \vert ^2 \rangle_t - \langle \vert \tilde{N}_\mathrm{~NGC1068}~(u,v) \vert ^2 \rangle_t \nonumber
 \label{eq5}
\end{eqnarray}
where $\tilde{I}_\mathrm{~NGC1068}$, $\tilde{I}_\mathrm{~calib}$,
$\tilde{N}_\mathrm{~NGC1068}$, and
$\tilde{N}_\mathrm{~calib}$ are
the Fourier transforms of the intensity distribution on individual
VISIR \textit{BURST mode} images of NGC~1068, of the calibrator
star, and of the background photon noise on the related images,
respectively. The final calibrated visibility estimate is

\begin{equation}
 \vert V(u,v) \vert ^2 = \frac{ \vert V_\mathrm{~NGC1068}~(u,v) \vert
 ^2 } {\vert V_\mathrm{~calib}~(u,v) \vert ^2}
 \label{eq6}
\end{equation}

In our study, the most important interest of \textit{BURST mode}
observations compared to long-time exposures is the ability to
calibrate the source spatial spectrum. We considered error propagation to determine error bars
for the uncalibrated visibilities. A Monte-Carlo computation is used
to estimate error bars for the calibrated visibility. It consists
in generating random numbers with a Gaussian distribution for $\vert
V_\mathrm{~NGC1068}\vert^2$ and $\vert V_\mathrm{~calib}\vert^2$
(as the histograms of the individual exposure 
estimates are Gaussians) taking their standard deviations into account. The standard deviation of the simulated visibility distribution is
the $1~\sigma$ error on $\vert V(u,v) \vert ^2$. Depending on spatial
frequencies, error bars on visibilities range between 1 and 10~\%.
Two-dimension (2D) maps of the visibilities and associated error bars
are obtained. Because the intensity distribution of the source is real, this 2D
visibility map is centro-symmetric with respect to the zero spatial
frequency (see left of Fig.~\ref{ellipses}). Cuts of the map along several
orientations are presented in Fig.~\ref{fits}. Orientations of the
spatial frequency plane correspond to the orientation on the sky plane
(conventions used being 0$\degr$ is related to the NS axis and 90$\degr$ to
the EW axis in images).






In the following, the visibilities are compared to simple geometrical
models as is often the case in optical-infrared interferometry. A
model of the spatial brightness distribution of the source
$O(\alpha,\delta)$ is used to extrapolate the high-spatial frequency
information from the measured low spatial frequencies. Thus, thanks to
the \textit{a priori} information injected through this model,
structures smaller than the diffraction limit of the instrument can
be estimated. This super-resolution technique works as long as the
signal-to-noise ratio on visibilities is high enough to be able to
detect a visibility drop with respect to 1.  In classical imaging, an
equivalent technique consists in detecting the width difference
between the point spread function (psf) and the source intensity
peak. This requires a high signal-to-noise ratio, however, and an
excellent stability of the psf.

\section{First study of VISIR visibilities} \label{study}

In this section, visibilities are compared to uniform disk models
that we used as rulers for measuring the typical sizes of structures in the object as a
function of azimuth. The complexity of models increases to account for the whole
evolution of the visibility and especially for the departure
from a circular symmetry.

  \subsection{Modeling} \label{modeling}

    \subsubsection{Azimuth-dependent modeling} \label{modeling_1}

\begin{figure*}[ht!]
  \begin{center}
   \includegraphics[height=4.5cm]{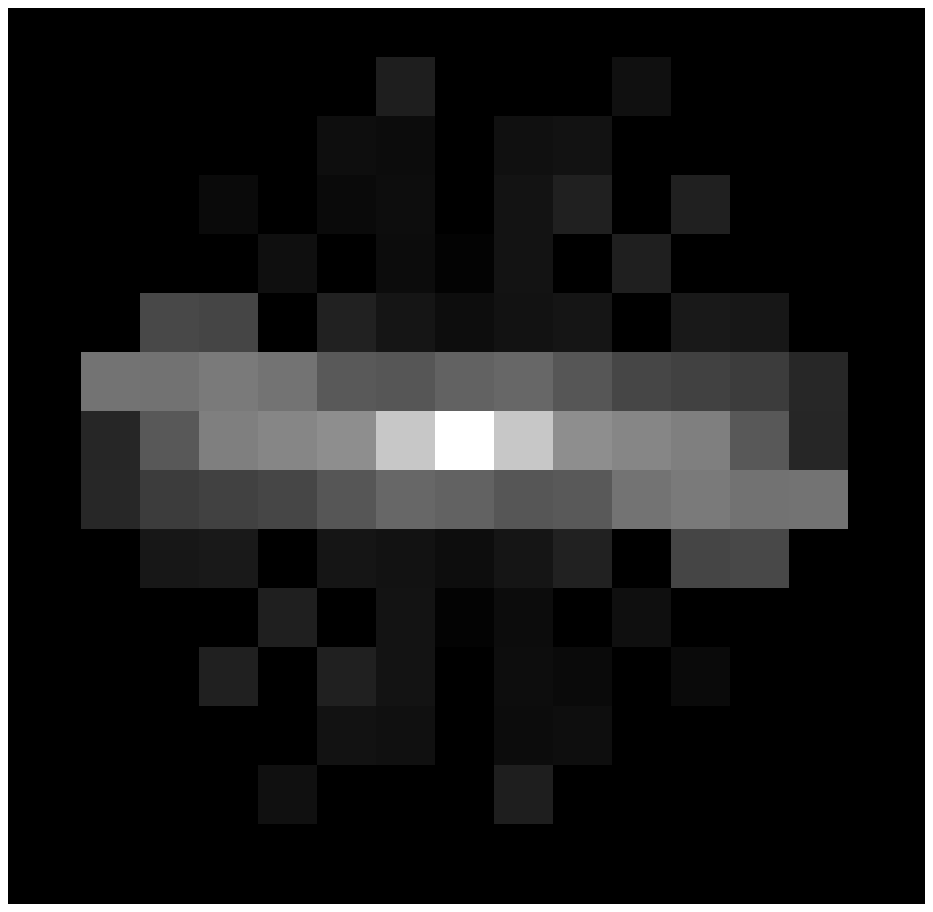}
   \includegraphics[height=4.5cm]{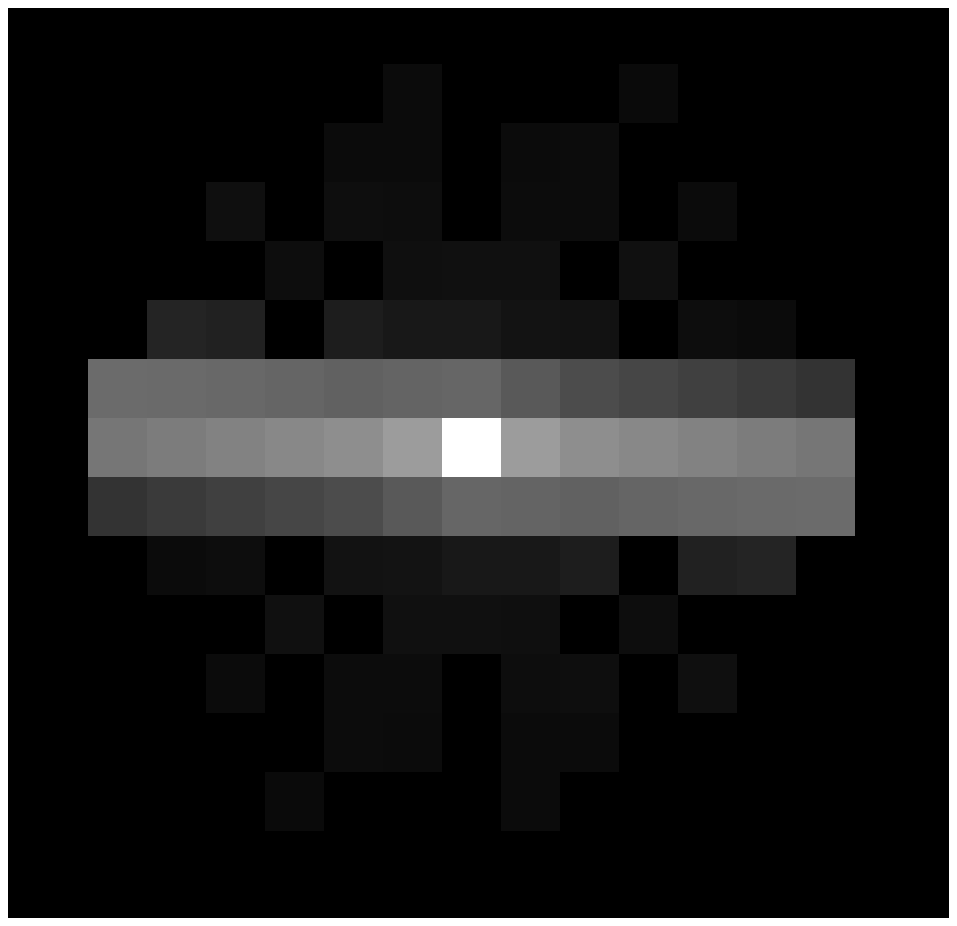}
   \includegraphics[height=4.5cm]{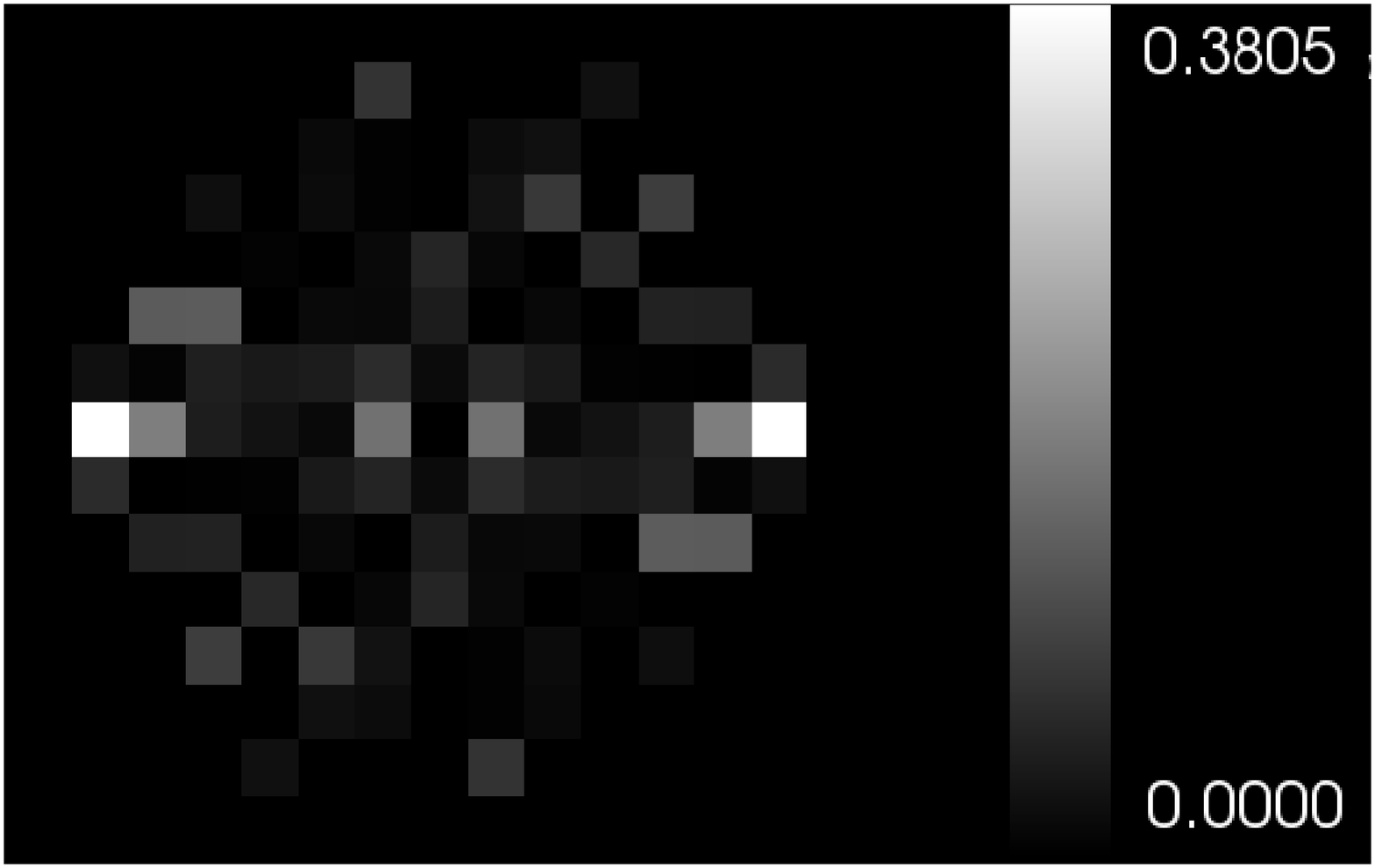}
  \end{center} 
 \caption{\textit{Left}:~visibility map as extracted from
 $BURST$ mode VISIR images. Only data points used for fits have value, others
 are set to 0. \textit{Middle}:~visibility map from the model of two
 circular disks and one
 elliptical uniform one (reduced $\chi^2$=8.7). These two maps
 are scaled between 0 and 1 (white points corresponding to a visibility value
 of 1 and black ones to value of 0). Comparison between left and right
 illustrates the quality of the fit. \textit{Right}: map of absolute
 terms of residuals, i.e. substraction of the modeled visibility map to the
 measurements one. The color scaling of the residuals map is presented
 on the right side.}
 \label{ellipses}
\end{figure*}

The 2D visibility map derived from VISIR images (see left of
Fig.\ref{ellipses}) shows an elongated shape in a direction close
to 90$\degr$. Similarly, cuts of this map along different
orientations (see Fig.~\ref{fits}) highlight the steep
fall in visibility along the NS axis (see plots corresponding to
0$\degr$ in Fig.~\ref{fits}) in comparison with the perpendicular axis (see plots corresponding to 90$\degr$). The
higher the visibility, the smaller the observed source. The source
is then well-resolved along the NS axis and smaller as we look
towards the perpendicular axis. A minimum of two uniform disks in the
model is necessary to fit the different cuts of visibilities. For each
orientation (marked by $\theta$), the free
parameters associated with the model are the sizes of the
two disks ($\oslash_1(\theta)$ and $\oslash_2(\theta)$) and the flux ratio between
them ($\eta(\theta)$). Fits are presented in Fig.~\ref{fits}. Along each direction, optimum parameters
were evaluated with a $\chi^2$ minimization and error bars on parameters
were obtained by varying the $\chi_\mathrm{min}^2$ by 1. The
global value of the reduced $\chi^2$ (i.e. corresponding to all fits)
is given by
\begin{eqnarray}
\chi^2_\mathrm{red} = \frac{1}{N-P_1}
    \sum_{i=1}^{N}\biggl[\frac{V_i^{2}-M_i~(~\oslash_1(\theta),\oslash_2(\theta),
    \eta(\theta)~)}{\sigma_i} \biggr]^{2}
 \label{chi2_1}
\end{eqnarray}
where ($N~-~P_1$) is the number of degrees of freedom, $N$
stands for the number of visibility points ($N$~=~91), and $P_1$ is the number of
free parameters. There are 3 parameters per orientation and a total of
12 orientations ($P_1$~=~36), $V_i$ and $\sigma_i$ are respectively
visibilities and error bars measured with the speckle analysis, $M_i$
are the modeled visibility values. The value of the global reduced
$\chi^2$ is 24.9.
The size of the inner component is $<$~300~mas, below the diffraction limit of the UT. The second component is well-resolved
($\sim$~1600~mas) and more or less circularly symmetric. For each component, sizes and error bars related to each orientation are
reported in the image and linked for the extended one (left panel in Fig.~\ref{fig3}). Nevertheless, it is difficult to figure out the real
shape of the components since part of the asymmetry is due to the
azimuth dependence of the flux ratio between the two components. It
ranges between $\sim$~0.29 along the EW axis and $\sim$~4.76 along
P.A.~=~14$\degr$ from N to E. A more realistic description of
MIR sources in the nucleus of NGC~1068 is given by the
global modeling explained below.

%

    \subsubsection{Global modeling} \label{modeling_2}

Using our previous results, we attempted to produce a global fit of
all orientations at the same time using a more complex model. The
global fit of the 2D map of VISIR was based on the most reliable visibility points (see left of 
Fig.\ref{ellipses}). We here consider the flux ratio between the components
to be a constant. To account for the asymmetry of the visibility
distribution, we first made use of a simple model of one circular, plus one
elliptical, uniform disk. This model converges to two different kinds
of solutions with an equivalent value of the reduced $\chi^2$=27. This
value is equivalent to the one obtained with the previous model
applied to each direction independently. It turns out that an additive
component was mandatory to have a better fit of the visibility
map and to better constrain the upper size limit of the smallest
component. To limit the number of parameters and account for the
asymmetry of the visibility map, this third model consists in two
circular uniform disks plus an elliptical one. There are seven free
parameters: the diameter of the two circular disks
($\oslash_1$ and $\oslash_2$); the semi-major and the semi-minor
axes $a$ and $b$, and the orientation $\theta$ of the elliptical disk
in the sky plane (angles are positive from north to east) ; the flux
ratios between the small and intermediate components ($\eta_1$) and
between the intermediate and the largest one ($\eta_2$).
\begin{eqnarray}
\chi^2_\mathrm{red} = \frac{1}{N-P_2}
    \sum_{i=1}^{N}\biggl[\frac{V_i^{2}-M_i~(~\oslash_1, \oslash_2, a,
    b, \theta, \eta_1, \eta_2~)}{\sigma_i}
    \biggr]^{2} 
 \label{chi2_2}
\end{eqnarray}
where $N$~=~91 and $P_2$~=~7.

\begin{figure*}[ht!]
  \begin{center}
   \includegraphics[height=5cm]{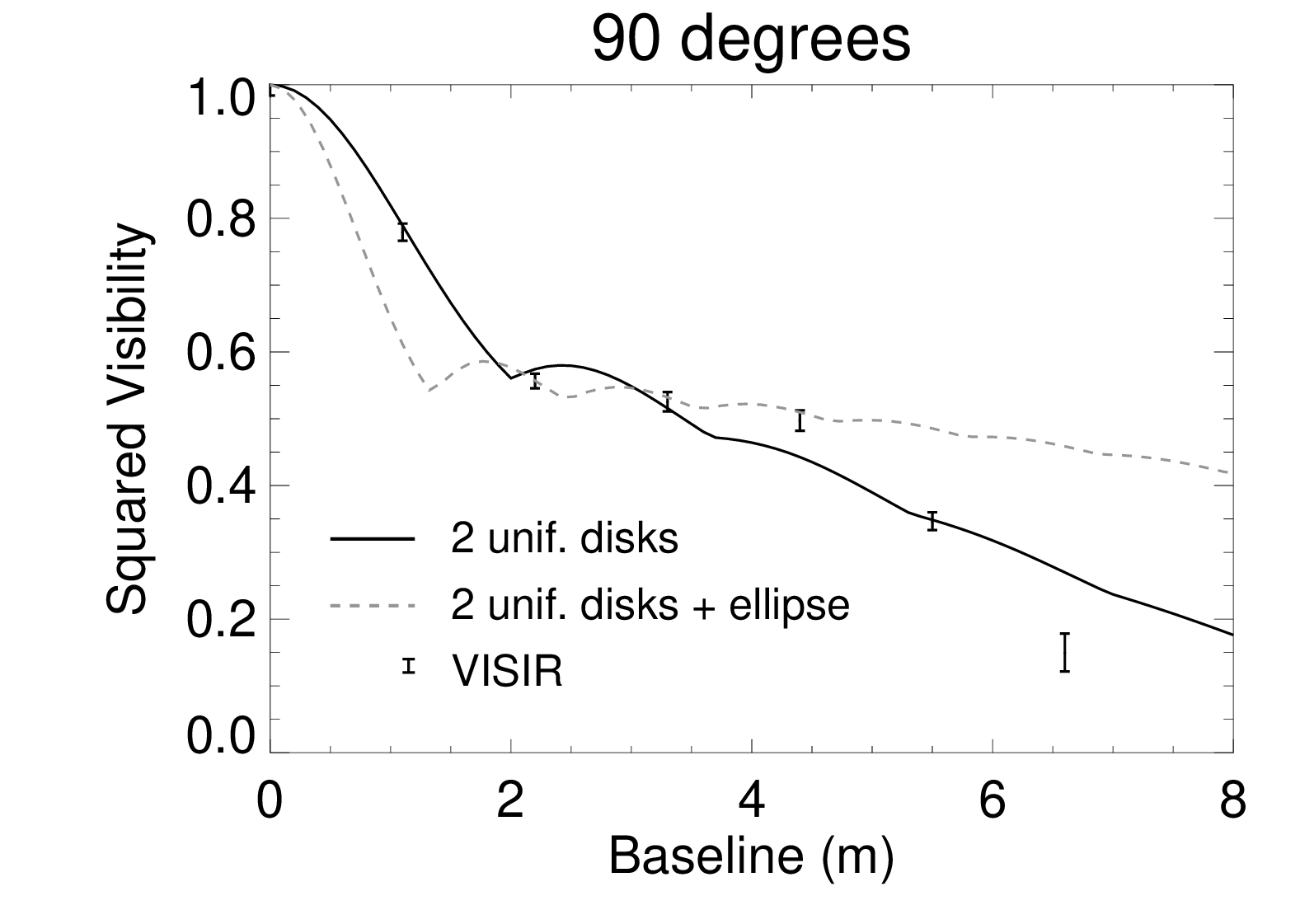}
   \includegraphics[height=5cm]{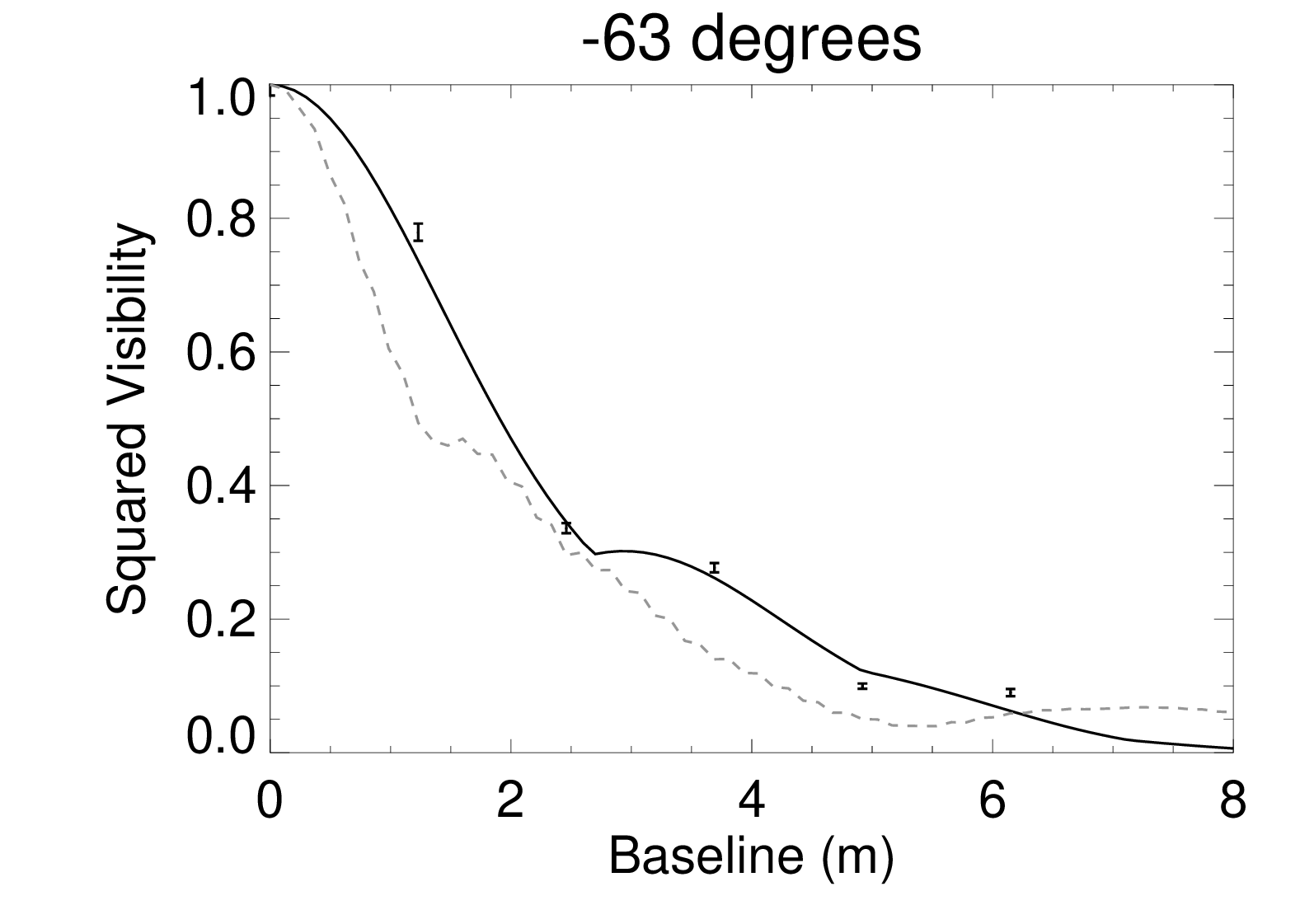}
   \includegraphics[height=5cm]{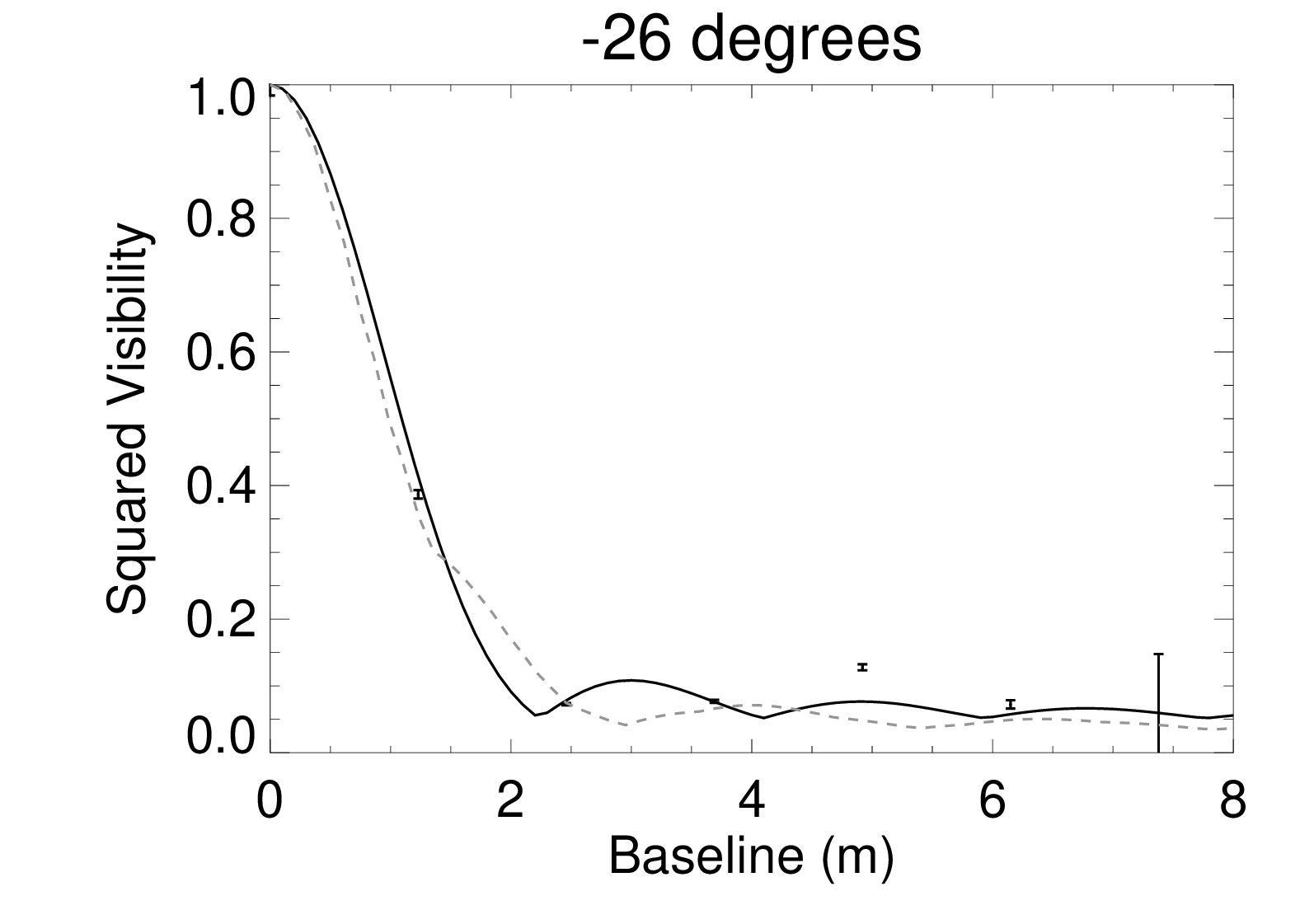}
   \includegraphics[height=5cm]{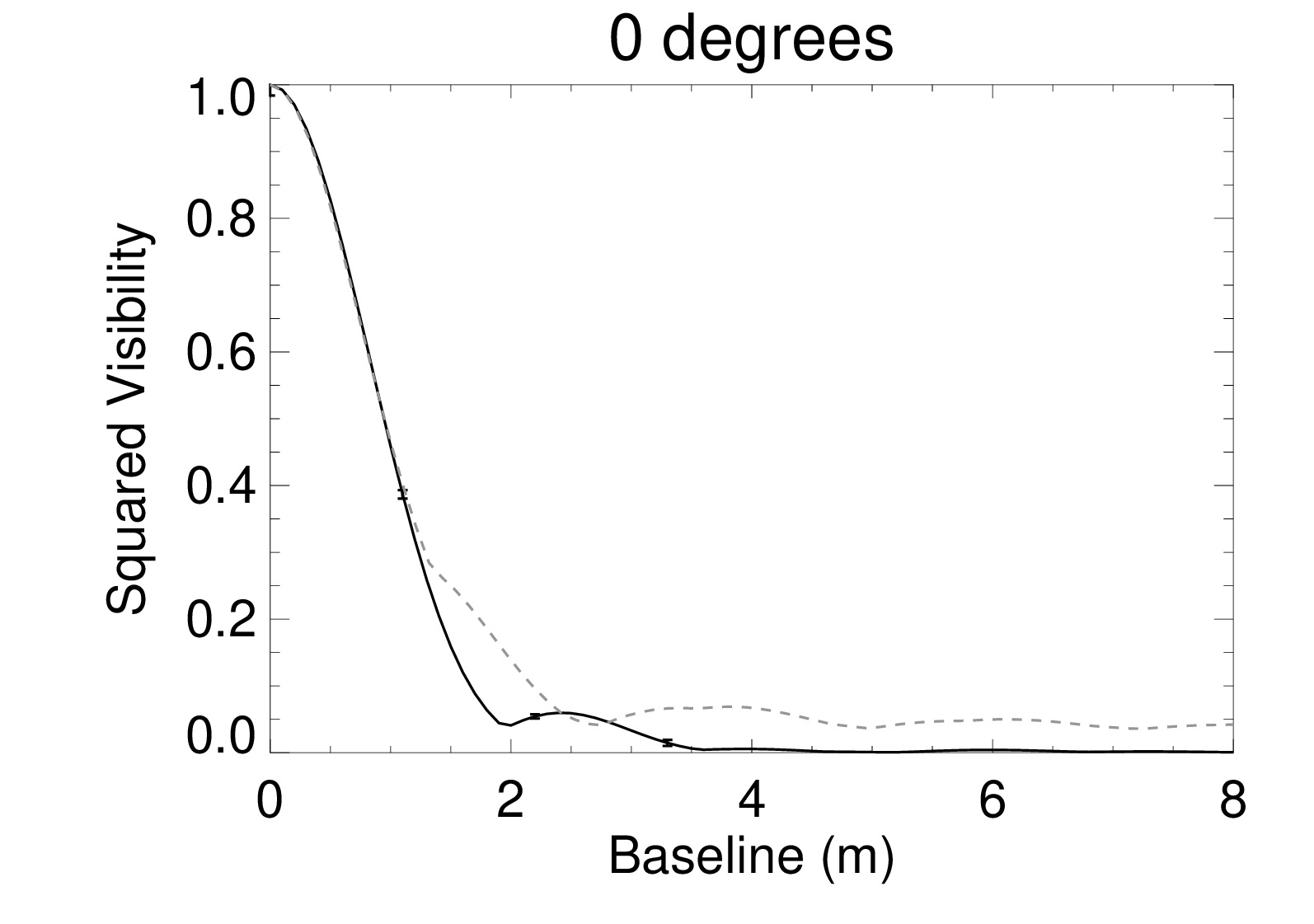}
   \includegraphics[height=5cm]{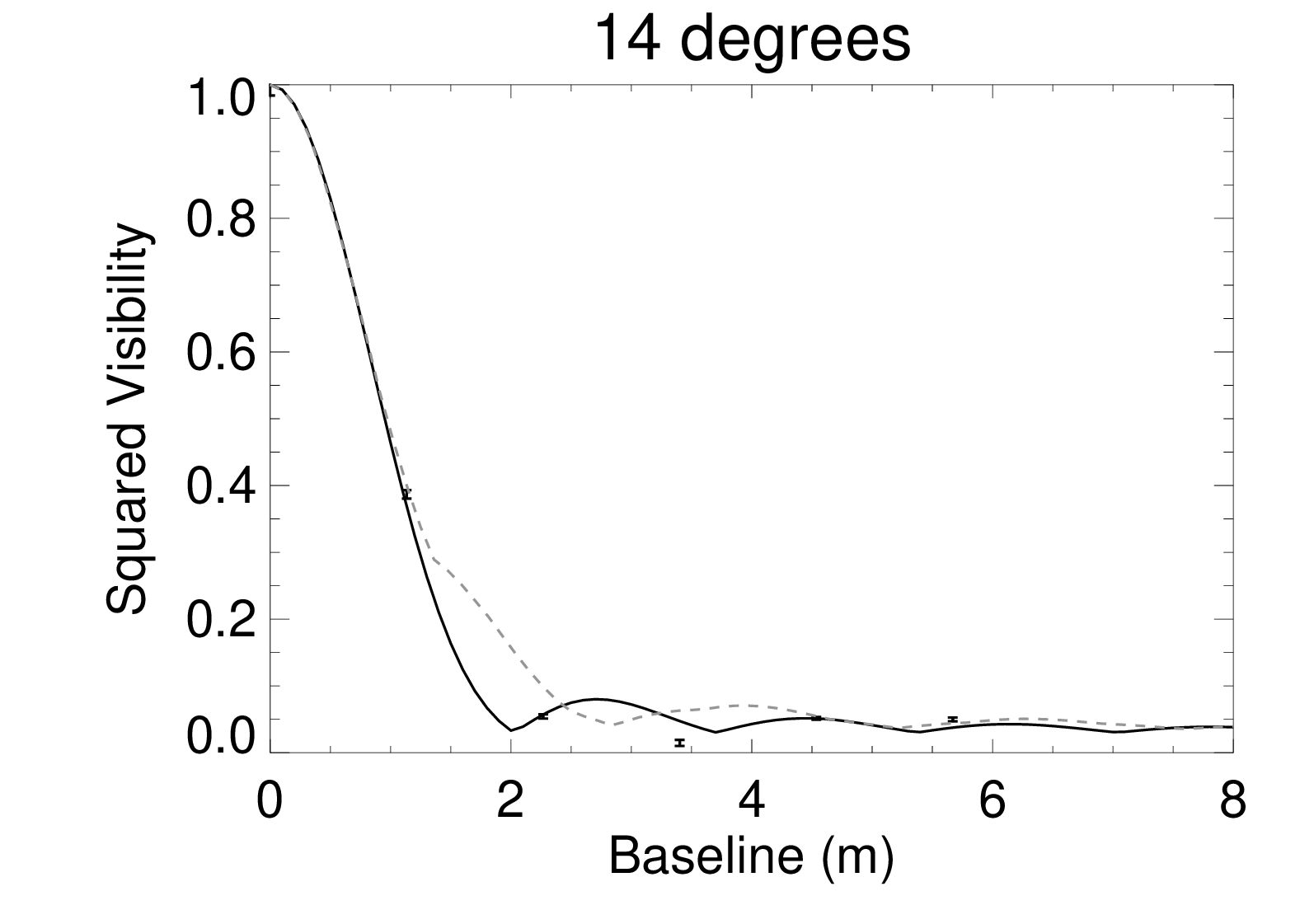}
   \includegraphics[height=5cm]{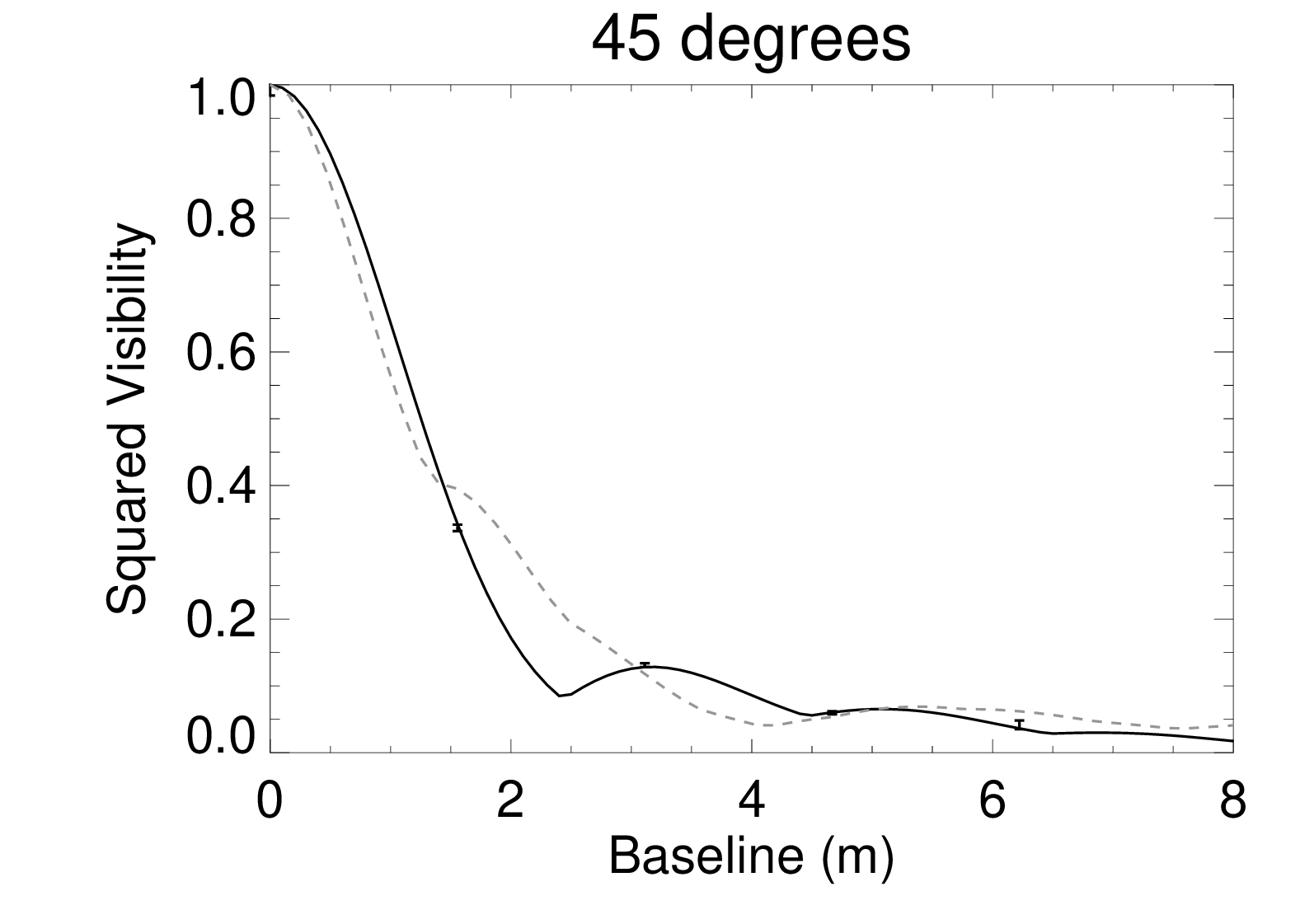}
   \includegraphics[height=5cm]{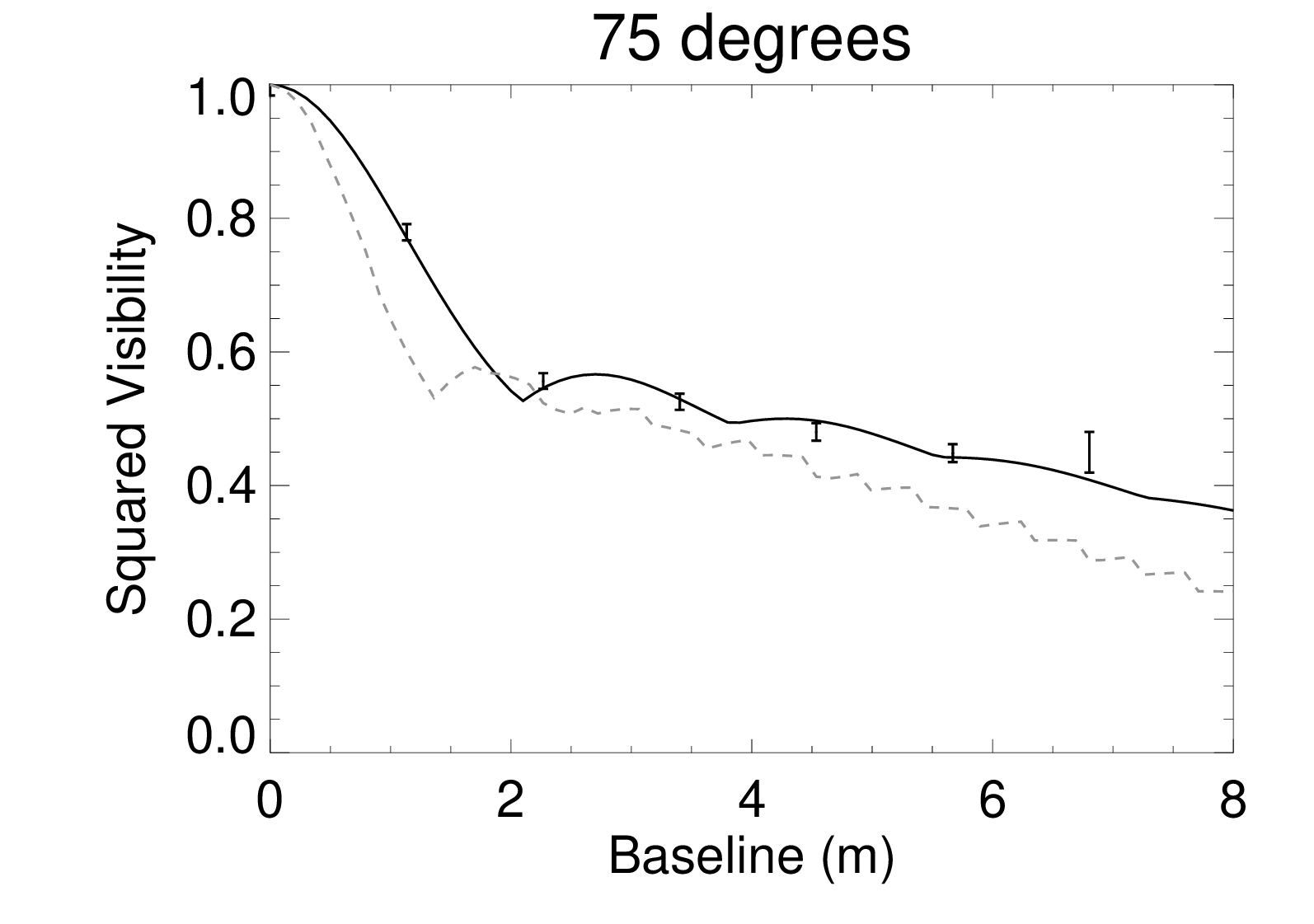}
  \end{center} 
 \caption{VISIR visibilities derived from \textit{BURST mode} images
 at 12.8~$\mu$m and associated error bars (symbols) along several
 directions (angles are positive from north to east). One can see the
 softer fall of the visibility toward the EW axis (90$\degr$) than
 toward the NS axis (0$\degr$), highlighting the NS elongation of the
 source. \textit{Black solid line} corresponds to fits with the two
 circular uniform disk model independently applied for each
 orientation (global $\chi^2_\mathrm{red}~\sim$~24.9). \textit{Dashed
 grey line} corresponds to the last model considered consisting in a
 two circularly and one elliptical uniform disk. This model is used in
 order to fit all baselines simultaneously. The value of the reduced
 $\chi^2$ is here 8.7. It is smaller than the value related to the
 previous model due to the higher number of freedom degrees of the
 global model (see Sect.~\ref{modeling}).}
 \label{fits}
\end{figure*}

The minimization of the previous expression of the reduced $\chi^2$
yields the following constraints on the parameters: \\
- $\oslash_1~<$~85~mas\\ 
- $a~<$~140~mas\\
- $b$~=~1187~$\pm$~22~mas\\
- $\theta$~=~-3.7$~\pm$~0.8$\degr$\\
- $\oslash_2$~=~2367~$\pm$~107~mas\\ 
- $\eta_1$~=~3.0~$\pm$~0.1\\
- $\eta_2$~=~1.4~$\pm$~0.1\\

Values for $\theta_1$ and $a$ are close to zero. We chose to
give an upper limit for these parameters as 0 is within the
$1\,\sigma$ range of uncertainty. The corresponding modeled
visibility and residuals maps are presented in the middle and on the
right of Fig.\ref{ellipses}, respectively. Fits
of visibilities by this model for different orientations in the
$(u,v)$ plane are compared to the ones obtained with the first model
of two circularly uniform disks in Fig.~\ref{fits}. The reduced
$\chi^2$ is equal to 8.7. It is smaller than with the first model for
which the number of degrees of freedom was much larger.

\begin{figure*}
 \begin{center}
  \includegraphics[height=6cm]{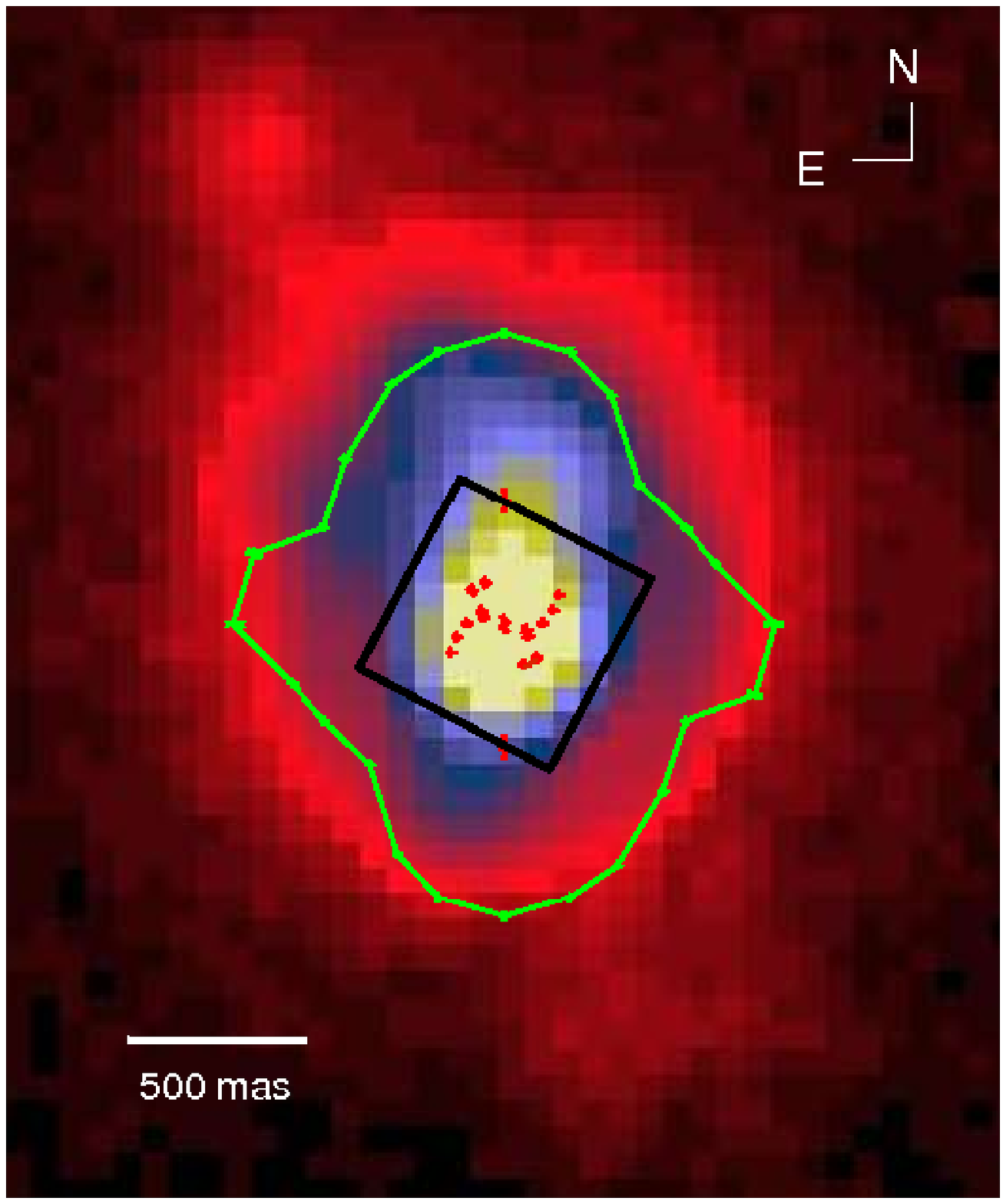}
  \includegraphics[height=6cm]{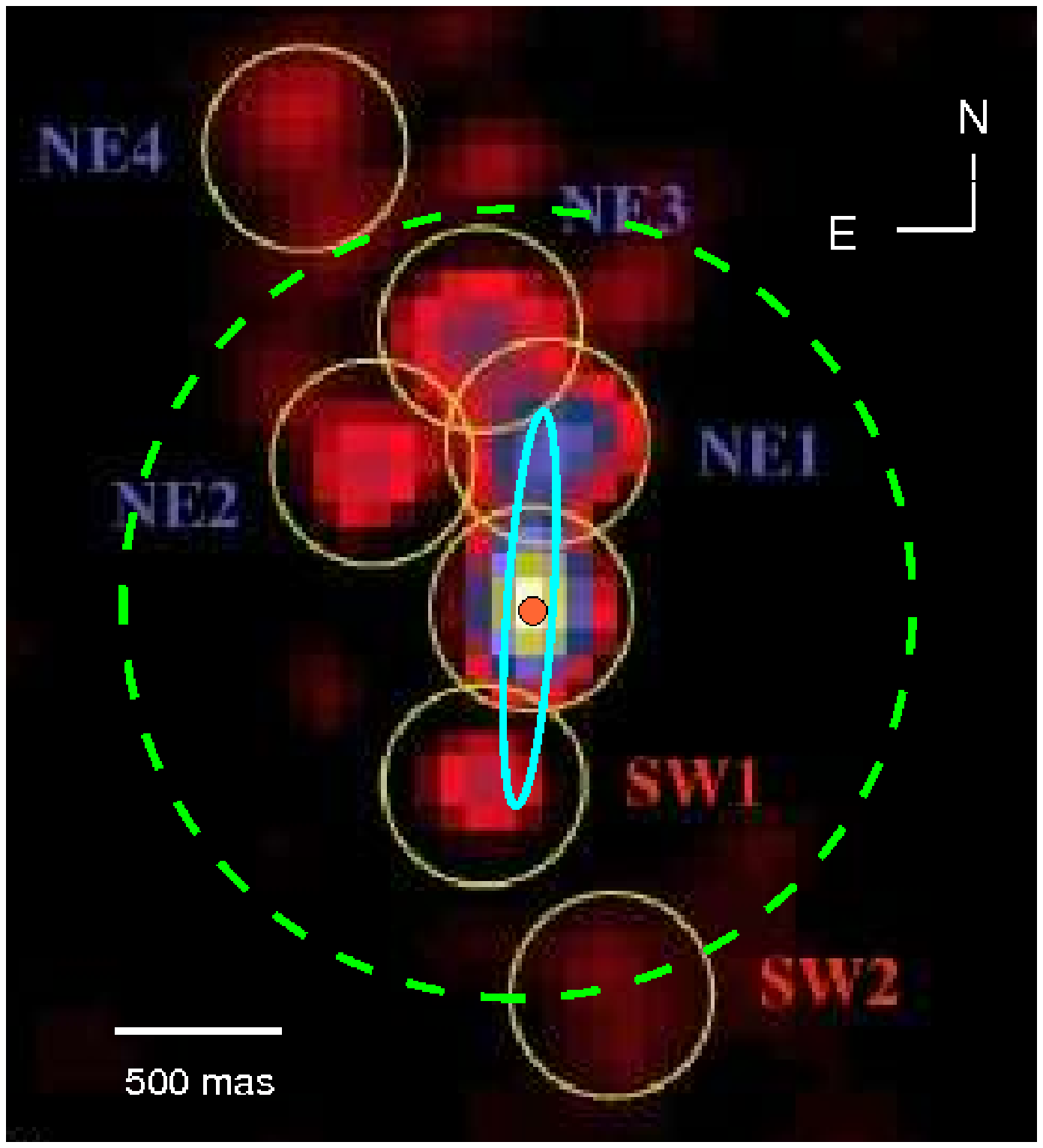}
  \includegraphics[height=6.4cm]{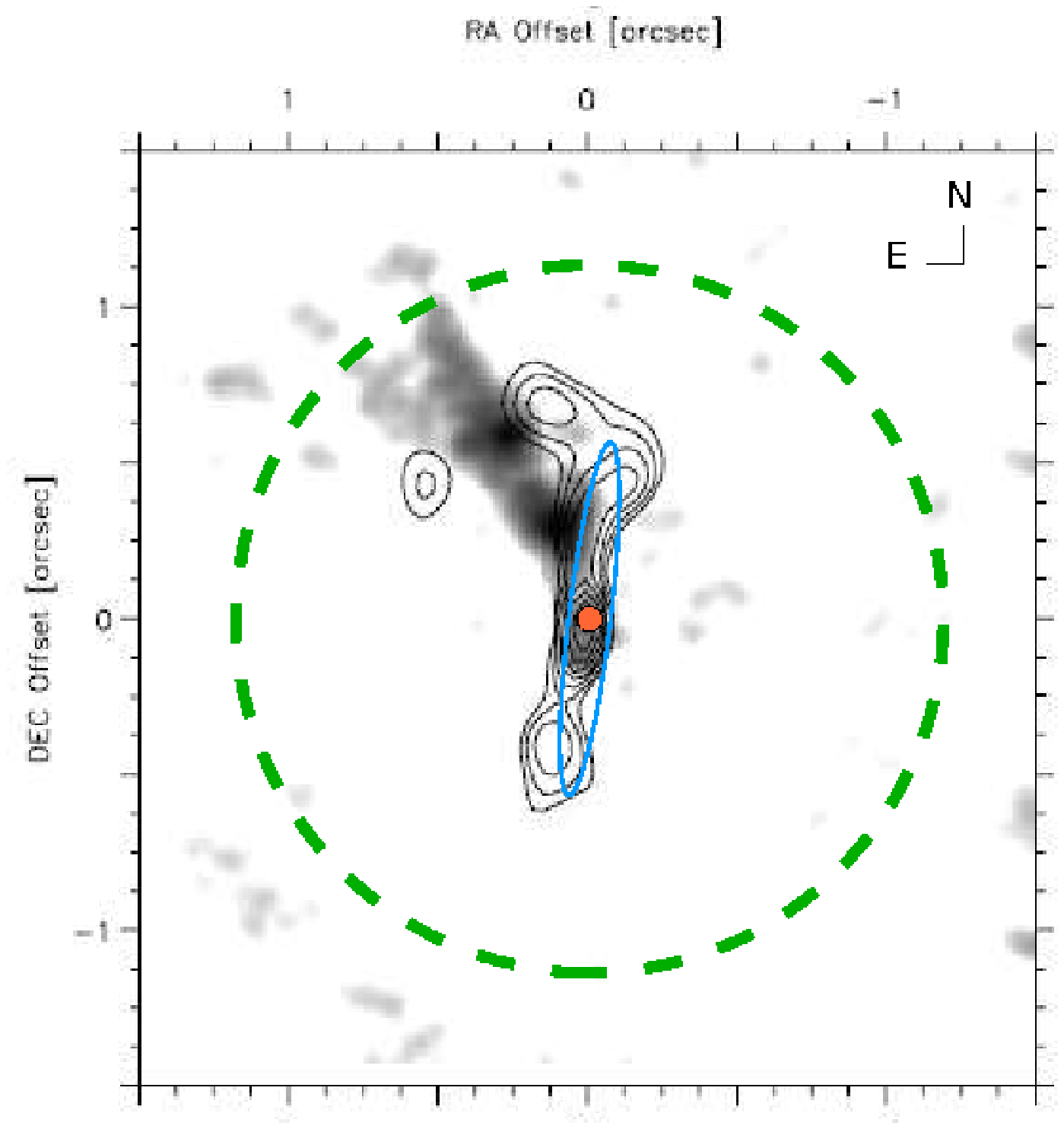}
 \end{center}
  \caption{ \textbf{Left:} comparison of different contributions to
  MIR emission in the core of NGC~1068. Background is the resulting
  12.8 $\mu$m VISIR \textit{BURST mode} image. Sizes derived from the
  two uniform disks model applied to each orientation and as
  described in Sect.~\ref{modeling_1} are superposed (\textit{red and
  green components}). \textit{The black square} displays the MIDI
  mask considered in the second part of the study of VISIR
  visibilities (see Sect.~\ref{comparison}). \textbf{Middle:}
  comparison between the structures resulting from the global
  modeling presented in Sect.~\ref{modeling_2} and the deconvolved
  image obtained from VISIR standard-mode observations \citep{Galliano_2005}. \textbf{Right:} same comparison
  with the contour plot of the 12.5~$\mu$m deconvolved image obtained
  at Keck 2 \citep{Bock_2000} and the 5GHz radio map of
  \citet{Gallimore_2004} (grey scale). \textit{Green dashed lines} correspond to
  the more extended
  component. \textit{The blue ellipse} corresponds to the elongated one
  at P.A.~$\sim$~-4$\degr$, of size
  ($<$~140)~mas~$\times$~1187~mas. \textit{The orange innermost component}
  represents the inner uniform disk $<$~85~mas. This component is
  directly associated to the \textit{dusty torus} resolved in
  interferometry with MIDI \citep{Jaffe_2004,Poncelet_2006}. }
  \label{fig3} 
\end{figure*}

For comparison with previous observations, geometrical parameters
constrained by this model are shown on the
deconvolved image of \citet{Galliano_2005} and \citet{Bock_2000} (see
the middle and right panels of Fig.~\ref{fig3}). The compact core and the ellipse account for the core and
the NS elongation \citep[i.e. knots NE1 and SW1 of][] {Galliano_2005}, while the 2$\farcs$4 uniform disk accounts (both
in size and total brightness) for the background of fainter structures
at larger scales. The comparison shows the consistency of the different processings
  applied to MIR images obtained at Keck~2 and with VISIR at the VLT.

\begin{figure*}
 \centering
 \includegraphics[width=16cm]{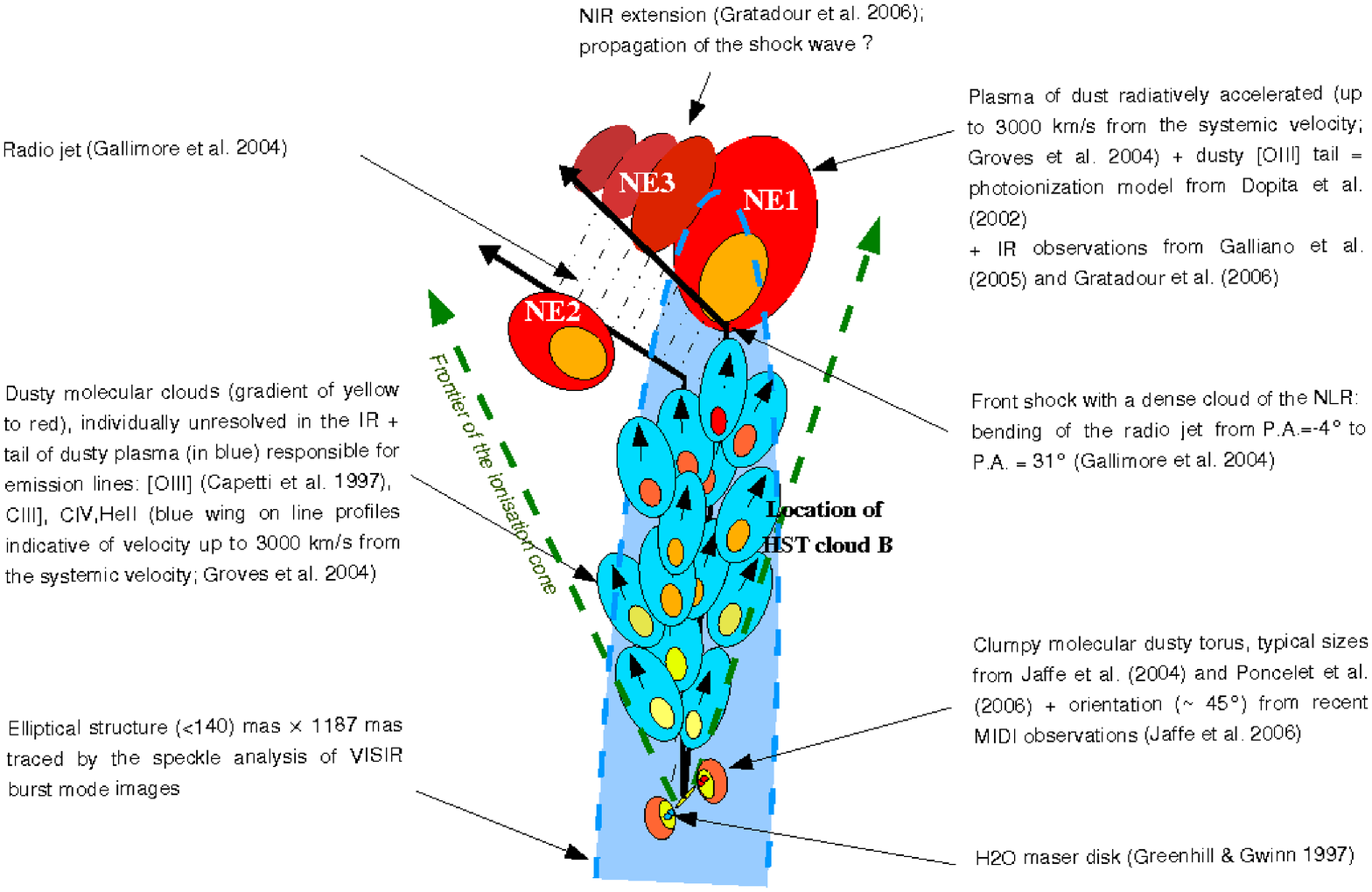}
  \caption{Picture summarizing the multi-wavelength structures observed toward the
  north in the nucleus of NGC~1068, as discussed in Sect.~\ref{elongation}. Among others, the representation
  of the radio jet (which bends from PA~=~-4$\degr$ to P.A.~=~31$\degr$) comes from observations of \citet{Gallimore_2004}, H$_2$0 masers from
  \citet{Greenhill_1997}, the \textit{dusty torus} from
  \citet{Jaffe_2004}, \citet{Poncelet_2006} and \citet{Jaffe_2006}, MIR clouds from the present study
  and from \citet{Galliano_2003}, NIR emission from
  \citet{Gratadour_2006}, [OIII] clouds from \citet{Capetti_1997},
  and UV lines from \citet{Groves_2004c}. The yellow color
  was chosen for heated dust, the gradient of yellow to red color
  is used for decreasing dust temperature (typically
  ranging between 300-500~K) and blue color for [OIII] UV clouds, for which HeII, CIII],
  and CIV emission-line profiles have a prononced blue wing indicative
  of proper motion up to -3000~km.s$^{-1}$, meaning that they are
  moving toward us. The blue elliptical structure derived from the
  analysis of VISIR \textit{BURST mode} images is discussed in Sect.~\ref{elongation}.}
  \label{global_picture} 
\end{figure*}

  \subsection{Discussion of the results} \label{discussion_modeling}

    \subsubsection{The MIR core} \label{core}

The speckle analysis provided a constraint in size, i.e. an upper limit of $\sim$~85~mas
for the inner source of emission in the nucleus. This is well below the
diffraction limit of the UT at 12.8~$\mu$m (i.e. 320~mas) or the FWHM
upper limits of 290 and 190~mas (along the NS and EW directions
respectively), which are measured from deconvolution of the 12.8~$\mu$m image
obtained with the standard acquisition mode of VISIR
\citep{Galliano_2005}.

This structure is directly associated with the \textit{dusty torus}
first resolved by interferometry using MIDI. It is in good agreement with
sizes produced by studies from both
\citet[][ FWHM~$<$~49~mas]{Jaffe_2004} and \citet[][ diameter $<$~82~mas]{Poncelet_2006}. The
present study does not allow any questions to be addressed about the orientation or inclination of this dust structure. However, evidence
from H$_2$O maser disks inclined at -45$\degr$ \citep{Greenhill_1997}
is thought to indicate the location of the inner edge of the \textit{dusty torus}
\citep{Kartje_1999}. This orientation has subsequently been confirmed by the analysis
of recent MIDI observations \citep{Jaffe_2006}, so we adopted it in Fig.~\ref{global_picture}.

In the NIR, the K-band core has been found to measure 18~mas~$\times$~39~mas
oriented at PA~$\sim$~16$\degr$ \citep{Weigelt_2004}. This
orientation differs from both the MIR and the distribution of
water masers. Likewise, deconvolved NACO images from
\citet{Gratadour_2006} trace structures $\sim$~30~mas and $<$~15~mas
in the K-band along the NS and EW axes, respectively. This K-band
structure is thought to be associated to the inner edge of the
\textit{dusty torus}, which is heated by radiation from the central engine.

    \subsubsection{The NS elongated structure} \label{elongation}
  
The ($<$~140)~mas~$\times$~1185~mas elliptical structure is aligned
with the base of the radio jet in the AGN of NGC~1068
(P.A.~$\sim$~-4$\degr$). The ellipse matches both the NE1 and SW1
knots of the deconvolved image of \citet[][ see middle
panel in Fig.~\ref{fig3}]{Galliano_2005}. The knot sizes are roughly equal
to the diffraction limit of the telescope so they must be understood as
upper size limits. In our case thanks to the super-resolution brought about
by the interferometric analysis technique, the elliptical disk is
thinner. Moreover, the
deconvolution process of \citet{Galliano_2005} may have emphasized the
intensity fluctuations in the structure just enough to produce the
knots. On this basis, our ellipse must be viewed as an average model of these
knots. Therefore, it is difficult to say which of the two
source representations is the best, but it should be noted that the superimposition of the
ellipse and contour plot from deconvolved 12.5~$\mu$m images from \citet{Bock_2000} is
very stricking (see right panel of Fig.~\ref{fig3}). Our model does
allow us to interpret a fraction of the flux of the central knot as part of
the elongated structure, whilest the remaining fraction belongs to the
central core (left unresolved by the image). Flux ratios
between the different components are discussed in Sect.~\ref{sec:flux}.

Examining the other wavelengths provides strong evidence for a correlation between the
NS extended core and HST [OIII] clouds A and B \citep[the location of
  HST cloud B is marked in Fig.~\ref{global_picture},][]{Capetti_1997}
located at 0$\farcs$1 S to N of the core, respectively. The NIR observations of \citet{Gratadour_2006} and fainter levels of
flux observed on the deconvolved images of \citet{Galliano_2005}
suggest no depletion of dust in the region between the core and knot NE1. 

The silicate feature evolution from the core up to
$\pm$~3$\arcsec$ outwards has been investigated by \citet{Rhee_2006} and
\citet{Mason_2006}. \citet{Rhee_2006} interpret this trend as dust emission
from the inner edge of the \textit{dusty torus}. If this is correct it
would mean that the torus should extend up to 1$''$ in the EW
direction, however neither interferometry nor VISIR observations have
shown this to be true. Indeed molecular dust has been found inside the
ionization cone up to 600~mas toward the N of the
central engine \citep{Mason_2006}. Thus drawing upon the conclusion of
\citet{Mason_2006} and using the theoretical work of \citet{Dopita_2002}, we explain our elliptical
structure and the correlation
between the NIR, MIR, and UV emissions in the nucleus of NGC~1068 in
terms of unresolved small clouds and of molecular dust irradiated and photoionized by
strong X/UV radiation fields produced by the central engine. Dust at
the front is directly heated by X-rays then photoevaporates and generates
a hot plasma of dust. This in turn is 
radiatively accelerated away from the central engine and emits in
both NIR and MIR. In addition an envelope of ionized gas is
produced that would account for the
narrow emission lines in the UV and optical ranges. It would
also explain the distribution of [OIII] clouds. This interpretation is strengthened by HST
spectroscopy \citep{Groves_2004c} that shows that the CIII], CIV, and HeII UV
line profiles have a pronounced blue wing within this region. This is
associated with the proper motions of velocities up to 3000
km.s$^{-1}$, which are indicative of NLR clouds being pushed toward
us. Fits of these line ratios point towards photoionization as the
dominating mechanism in comparison with excitation by
radio-jet/interstellar medium shocks. Thus, NLR dusty clouds observed
in NIR, MIR, UV must be at the front of
the ionization cone and accelerated towards us in a radiative wind (see
Fig.~\ref{global_picture}). The thinness of the ellipse can then be
explained either by the ionization cone
having a very small opening angle $\lesssim$~50$\degr$ (i.e. the angle
between the two frontiers of the ionization cone), meaning that the
opening angle of the torus would be $\gtrsim$~150$\degr$ (i.e. the
angle between nothern and southern walls of the torus) or by the
motion of the small unresolved dusty clouds being frozen in that of the
radio jet. In both cases, their propagation would have to be confined to a
thin region, such as what is observed.

The effects of jet-NLR clouds interactions are weak on the UV
spectra \citep{Groves_2004c} so could not account for the high color
temperature observed in the NIR range so far away from the central engine
\citep{Gratadour_2006}. However, it is stricking that the elongated ellipse is aligned with the
radio jet close to the core. There is actually some evidence for local jet-NLR
interactions, such as the bending of the jet axis initially at
PA~=~-4$\degr$ to PA~=~31$\degr$ \citep{Gallimore_2004}. This
would indicate a direct collision between the radio jet and a dense NLR cloud --
labeled HST-C in UV \citep{Capetti_1997}, NE1 in MIR
\citep{Galliano_2005}, and IR-1b in the NIR \citep{Gratadour_2006}. A
twisting of structures involving these two PA is found in the
deconvolved MIR and NIR images
\citep{Galliano_2005,Gratadour_2006}. The peculiar distribution of
knots above cloud NE1 may be the signature of the shock-wave propagation
in the dusty plasma medium as already discussed by
\citet{Gratadour_2006}. The global picture of this scenario is given in
Fig.~\ref{global_picture}.

    \subsubsection{More extended structures} 

The extended component of the model of the VISIR visibilities has been
used to account for the most extended contributions to the
flux. Given the deconvolved image of \citet{Galliano_2005}, it
includes knots NE2, NE3, NE4 and SW2 (see the middle panel of
Fig.~\ref{fig3}). The geometry we considered is clearly too simple,
but a detailed description of the extended environment of the core of
NGC~1068 goes beyond the scope of the present study. The important
information to consider here is the flux ratios between the total flux
from the extended environment and from the inner structures.

     \subsubsection{Flux ratios between components}\label{sec:flux}

We compare the flux ratios values from our global modeling, equal to 3 and 1.4
between the different components, with observations
of \citet{Bock_2000} and \citet{Galliano_2005}. First, our inner
component is associated to the source of flux labeled $b$ in
Fig.~5 of \citet{Bock_2000} and the ellipse (referred to {\it elongated} component in the
following) to those labeled $a$ and $c$. According to their mean fluxes, we deduce
$F_{\mathrm{core}}/F_{\mathrm{elongated}}~=~2.92$.  This ratio is in
very good agreement with our best fit.

Then, using Table~1 of \citet{Galliano_2005} and associating both our
 elongated component with the SW1 and NE1 knots and our extended
 uniform disk with the remaining SW and NE knots, we have computed the
 following flux ratios: 
\begin{itemize}
\item  $F_{\mathrm{core}}/F_{\mathrm{elongated}}~=~2.02 $
\item $F_{\mathrm{elongated}}/F_{\mathrm{extended}}~=~1.72$.

\end{itemize}

Given the uncertainties on photometry measurements in deconvolved
images and the simplicity of our model, and
considering that the structures of \citet{Galliano_2005} are large in
size compared to the diffraction limit of the telescope, these ratios
are in good agreement with those derived from our
study. Since the elongated structure in the EW direction and the central core are
not resolved by the telescope, both fluxes are biased, and this can
explain part of the discrepancy between the two estimates.

We conclude that our simple interferometric model has captured most of
both the sizes and relative intensities of the object structures when
compared to the deconvolved images. It offers extra information thanks
to the higher resolution achieved, both on the core associated to the
dusty torus and on the width of the NS elongated structure.

\section{Comparison with MIDI observations}\label{comparison}

\subsection{How to compare VISIR and MIDI visibilities ?} 

One of the important aims of the study is the link between low spatial-frequency visibilities derived from VISIR images, with long baselines
data points obtained with MIDI and the radiative
transfer model by \citet{Poncelet_2006}.  Visibilities extracted in
part \ref{observations} correspond to a field of view for VISIR of
several arcseconds, whilest the field of view for MIDI is limited to a
slit of 0$\farcs$6$\times$~2$''$ oriented at -~30$\degr$. In addition, the emerging flux is mainly limited to what comes from the
core by a $\sim$~0$\farcs$6$\times$~0$\farcs$6 mask oriented at
-~30$\degr$ centered on the nucleus (left panel of Fig.~\ref{fig3}).

In processing images such as the conditions of MIDI observations we first
unbias spectra from the pixelization of VISIR images, and secondly we
take the mask of MIDI into account. To correctly
position the mask on the flux maximum, we have to keep the spatial
information of images contained in the phase of spectra (in addition of
the unbiased moduli). New spatial spectra of NGC~1068 and the
calibrator are then calculated via the following equations:

\begin{equation}
 S_\mathrm{NGC1068}(u,v) = \vert V_\mathrm{~NGC1068}~(u,v) \vert~\times~e^{-i~\Phi_\mathrm{NGC1068}}
\end{equation}

\begin{equation}
 S_\mathrm{calib}(u,v) = \vert V_\mathrm{calib}~(u,v) \vert~\times~e^{-i~\Phi_\mathrm{calib}}
\end{equation}
where $\vert V_\mathrm{~NGC1068}~(u,v) \vert$ and $\vert
V_\mathrm{calib}~(u,v) \vert$, $\Phi_\mathrm{NGC1068}$ and
$\Phi_\mathrm{calib}$ are respectively the moduli unbiased from noise
(as calculated in Sect.~\ref{observations}) and the phases of the
spatial spectra of NGC~1068 and of the calibrator star.
  
The bias produced from pixelization effects is corrected by dividing moduli by
the pixel transfer function in the Fourier plane, the pixel being considered
as a simple filter with a width of 75~mas (i.e. the size of one pixel) in the
image plane. For oversampling, original 32~$\times$~32 maps are padded out
with zeros after the cut-off frequency of the UT (i.e. $D/\lambda$),
to achieve 256~$\times$~256 maps. The resulting maps produce new
pixels angular size of $\sim$~9.4~mas.pixels$^{-1}$. The MIDI mask coded onto a
256~$\times$~256 image is Fourier transformed. Spectra of NGC~1068
and of the calibrator star are convolved with the spectrum of the
mask, written as

\begin{equation}
 S_\mathrm{mask}(u,v) = \vert V_\mathrm{mask}~(u,v) \vert~\times~e^{-i~\Phi_\mathrm{mask}}
\end{equation}

Low spatial-frequency visibilities derived from
VISIR images as obtained in the same conditions of MIDI data
acquierement are given by

\begin{equation}
 \vert V(u,v) \vert = \frac{\vert S_\mathrm{NGC1068}(u,v) *
 S_\mathrm{mask}(u,v) \vert}{\vert S_\mathrm{calib}(u,v) * S_\mathrm{mask}(u,v) \vert}
\end{equation}
where * stands for the convolution.

MIDI observations made in 2003 correspond to baselines
oriented at 0$\degr$ and -45$\degr$. Therefore only cuts along
these two orientations of low spatial-frequency visibility maps
have been retained. For individual exposures, signal-to-noise ratio
is low and phases are noisy. Therefore error bars on visibilities are not computed from the statistical
distribution of individual
measurements as in Sect.~\ref{observations}, but from the reduced
$\chi^2$ of fits at 0$\degr$ (see Fig.~\ref{fig6}), which has been
forced to be equal to 1. It is equivalent to considering the model as correct and to
interpreting discrepancies with the data as errors. This leads to a
qualitative estimate of the error of $\sim$~0.026, which has been applied to all visibilities between 0 and 8~m.
The comparison with MIDI data points is presented in Fig.~\ref{fig5}.

 \subsection{The link between MIDI and VISIR observations}\label{link}

\begin{figure}
 \centering
 \includegraphics[width=9cm]{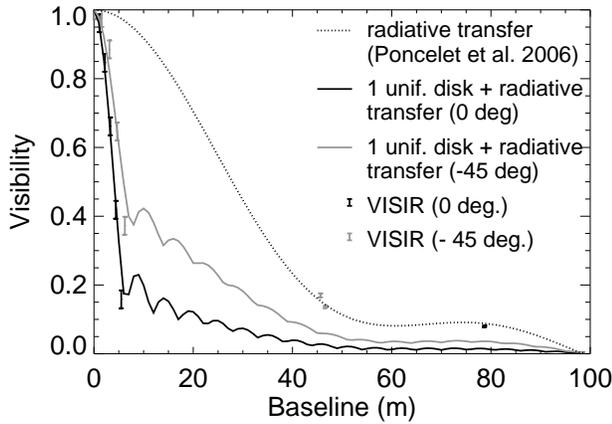}
  \caption{Comparison between high spatial-frequency visibilities from
  MIDI at 12.8~$\mu$m (data points at 45 and 78~m) and low spatial frequency ones derived from
  VISIR \textit{BURST mode} images including the mask discussed in
  Sect.~\ref{comparison} (points between 0 and 8~m of baseline). The
model of the two uniform disks (\textit{black and grey solid lines}) has to be compared with the radiative transfer
  model of \citet{Poncelet_2006} applied to MIDI data (\textit{black
  dotted line}). A third component has been added to the
  one already revealed by the radiative transfer, to partially account
  for the steep fall of the visibility at short baseline. Size of this component is $\sim$~600 mas
  and $\sim$~300~mas along directions oriented at 0$\degr$ and -45$\degr$ respectively.}
  \label{fig5} 
\end{figure}

 Figure~\ref{fig5} is the comparison between $a)$ low spatial-frequency
 visibilities from VISIR images considering the MIDI mask, $b)$ high
 spatial frequency ones from MIDI, and $c)$ the model of
 \citet{Poncelet_2006} that reproduces MIDI data. The
 unexpected sharp fall of low spatial frequencies
 visibilities demonstrated how constraining these measurements are for modeling long baseline interferometric data.

 On the one hand, short baselines visibilities are well-reproduced by
 a uniform disk model of sizes $\sim$~450~mas and $\sim$~350~mas along
 directions at 0$\degr$ and -45$\degr$, respectively. On the other
 hand, high spatial-frequency visibilities from MIDI are represented
 by a radiative transfer model inside a dusty layer ranging in size
 between 35~mas and 82~mas \citep{Poncelet_2006}. The initial attempt to fit
 both sets of data simultaneously was associated with the model of the dusty layer of
 \citet{Poncelet_2006} being surrounded by a uniform disk. The only free parameters were the size of the
 uniform disk and the flux ratio between the disk and the dusty
 layer. This model agrees with the strong fall of visibility at low
 projected baselines but fails to account for the faint level of
 visibility at high projected baselines (see Fig.~\ref{fig5}). This
 forces us to attempt a new fit in which we have to consider the parameters of
 the dusty layer of \citet{Poncelet_2006} as free. This is
 not yet possible according to the limited amount of \textit{BURST
 mode} data sets that are currently available, as the wider
 spectral range that is required.

\begin{figure}
 \begin{center}
 \includegraphics[width=9cm]{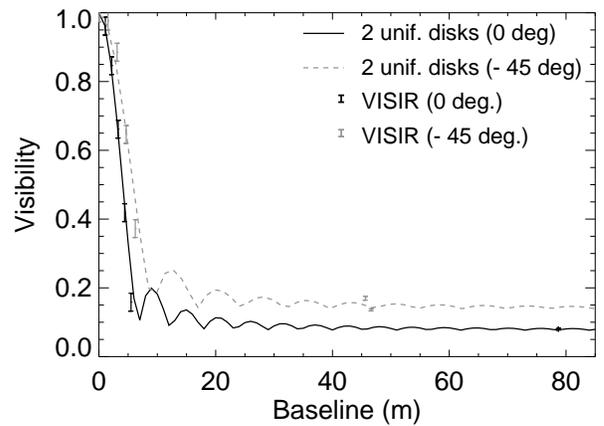}
 \end{center}
  \caption{Comparison between high spatial-frequency visibilities from
  MIDI at 12.8~$\mu$m and low spatial frequency ones derived from
  VISIR $BURST mode$ images including the MIDI mask as discussed in
  Sect.~\ref{comparison}. \textit{Solid black and dashed grey lines}
  correspond to a model of two circularly uniform disks applied along
  both directions at 0$\degr$ and -45$\degr$ independently. Optimum parameters
  are: $\oslash_1(0\degr)$ and
  $\oslash_1(-45\degr)~<$~21~mas,
  $\oslash_2(0\degr)$~=~479~$\pm$17~mas,
  $\oslash_2(-4\degr)$~=~450~$\pm$~9mas,
  $\eta(0\deg)$~=~12~$\pm$~5 and $\eta(-45\degr)$~=~6~$\pm$~2.}
  \label{fig6} 
\end{figure}

Nevertheless, to be able to quantify the fraction of extended emission entering the field of view of MIDI, compared to the emission of the compact
 core, we fit both low and high spatial-frequency visibilities together
 in both directions using a simple model of two uniform disks, in
 which free parameters are the sizes and the flux ratios between the two disks. The
 minimum value of the reduced $\chi^2$ is 3.42. This corresponds to an
 unresolved inner disk of upper size limit $\sim$~21~mas in each
 direction. The extended component has a size of 479~$\pm$~17~mas
 along the NS direction and $\sim$~350~$\pm$~9~mas along
 PA~=~-45$\degr$. The flux ratios between the two components are
 found to be 12~$\pm$~5 and 6~$\pm$~2 in these two directions (see
 fits in Fig.~\ref{fig6}).


 \subsection{Discussion}

This study demonstrates that there are two main
structures appearing in the MIDI field of view, a compact one of size $\sim$~20~mas 
surrounded by an extended and slightly elongated one (the size
ratio between -45$\degr$ and 0$\degr$ is $\sim$~27\% for this
second component). The first structure is associated to the dusty layer
highlighted solely by the MIDI data \citep{Jaffe_2004,Poncelet_2006}. Sizes
are slightly different due to the fact that it is a more simplistic
model than those of \citet{Jaffe_2004} and \citet{Poncelet_2006}. In
addition, the direct environment surrounding the compact core (of size
$\sim$~500~mas) that enters the field of view of MIDI was
not taken into account by previous studies. According to the flux ratio between components,
it contributes to more than 83~\% of the total flux seen by MIDI.

Constraints on temperatures of the \textit{dusty torus} derived from only
MIDI data \citep{Poncelet_2006} are not biased by this
effect. Indeed they come from an additional information being the
spectral energy distribution supplied by MIDI. They are found to be on the order
of $\sim$~ 300~K, a value supported by the continuum slope of
the central 0.4$''$ N-band spectra \citep{Mason_2006}. That
temperatures are so low so close to the central engine is strong
evidence of the clumpy nature of the \textit{dusty
torus} \citep{Elitzur_2006}.

The compact central component is already traced with the study of full
field images presented in Sect.~\ref{study}: it is directly associated
to the innermost component $<$~85~mas. The extended component, of
sizes 479~$\pm$~17~mas and 350~$\pm$~9~mas along PA~=~0$\degr$
and -45$\degr$, is not entirely related to the
elliptical NS elongated ellipse traced in Sect.~\ref{modeling_2} since
this last one does not completely enter in the field of view of MIDI. However, according to the discussion
in Sect.~\ref{discussion_modeling} and to the photoionization
scenario of \citet{Dopita_2002}, this elongated component could be associated to
small dusty clouds heated by the radiation field from the central
engine and photoevaporating. The NIR and UV counterparts of these clouds
\citep{Gratadour_2006,Capetti_1997,Groves_2004c} and the evolution of
the 9.7~$\mu$m silicate emission toward the north of the core
\citep{Mason_2006} is evidence for the presence of dust inside the
ionization cone at distances less than 300~mas from
the central engine (i.e. half of the MIDI mask). The present study
supports this fact and demonstrates the need to consider the
close environment of the \textit{dusty torus} for modeling
interferometric data. 

The small set of MIDI and VISIR
BURST mode data currently available only allows the use of simple
geometrical models for comparing low and high spatial-frequency
visibilities. It is strongly necessary to apply this approach at different wavelengths to get a better description of the
source of mid-IR emission seen by MIDI, associated with the dusty
torus, and linked with its immediate surroundings.

\section{Conclusions} \label{conclusion}

The speckle analysis of the 12.8 $\mu$m \textit{BURST mode} VISIR images
constrains the upper size limit of the MIR core of NGC~1068 to
85~mas. This is well below the diffraction limit of the UT which
demonstrates the
\textit{achievement} of the \textit{BURST mode} of VISIR. The NS elongated
component is described here as an elliptical structure oriented
at PAs~=~-4$\degr$ (such as the base of the
radio jet) and of size ($<$~140)~mas~$\times$~1187~mas. Following the
scenario proposed by \citet{Dopita_2002}, it is interpreted as small
dusty clouds individually unresolved, distributed inside the
ionization cone, and photoevaporating. This explains the correlation of
observations at different wavelengths. 

The use of VISIR \textit{BURST mode} images for the comparison with MIDI data reveals the strong and
unexpected fall in visibility between 0 and 6~m of baseline. The
use of simple geometrical models accounts for the two data sets
simultaneously and distinguishes between two main components: a
compact one $\sim$~20~mas associated to the \textit{dusty torus}
traced by interferometry with MIDI \citep{Jaffe_2004,Poncelet_2006}
and an extended one ($<$~500~mas) entering in the field of view of
MIDI. This surrounding environment, partially related to the elongated
elliptical structure, contributes to more than $\sim$~83$\%$ of the
flux emitted by the core. It is therefore mandatory to take it into account when modeling the MIDI data. By doing this we do not challenge the
low temperatures of the \textit{dusty torus} constrained by the
previous analysis of MIDI data. Besides, it reinforces the scenarios of clumpy torus.

According to the small set of VISIR \textit{BURST mode} and MIDI data available
up to now, we reach the limits of simple descriptions of sources of
mid-IR emission in the AGN of NGC~1068. To further describe the
complexity of the compact dusty core and the deeper link with its
direct surroundings, the same processing of
VISIR \textit{BURST mode} images has to be performed at other wavelengths. This will be possible
when the BURST imaging mode of VISIR is open for observations at
the VLT.  Moreover, as it shows how constraining low spatial frequency
visibility points are, this study underlines the need to further fill
the gap of interferometric data between 8 and 45~m of projected
baseline. On the one hand, baselines between 30 and 45~m are reachable with the VLTI
and the UTs. On the other, since most of the ATs are now installed at Paranal,
it will be interesting to attempt observations with these 1.8~m pupil
telescopes to bridge the gap from 8 to 23~m to allow for more accurate multi-scale modeling 
of the inner part of NGC~1068.

\begin{acknowledgements} 
We would like to thank Eric Thiebaut, Ferreol Soualez, and Renaud Foy for
their help during the treatment of VISIR images.

\end{acknowledgements}

\bibliographystyle{aa}
\bibliography{poncelet_references}

\begin{thebibliography}{24}
\expandafter\ifx\csname natexlab\endcsname\relax\def\natexlab#1{#1}\fi

\bibitem[{{Antonucci} \& {Miller}(1985)}]{Antonucci_1985}
{Antonucci}, R.~R.~J. \& {Miller}, J.~S. 1985, \apj, 297, 621

\bibitem[{{Bock} {et~al.}(2000){Bock}, {Neugebauer}, {Matthews}, {Soifer},
  {Becklin}, {Ressler}, {Marsh}, {Werner}, {Egami}, \& {Blandford}}]{Bock_2000}
{Bock}, J.~J., {Neugebauer}, G., {Matthews}, K., {et~al.} 2000, \aj, 120, 2904

\bibitem[{{Capetti} {et~al.}(1997){Capetti}, {Macchetto}, \&
  {Lattanzi}}]{Capetti_1997}
{Capetti}, A., {Macchetto}, F.~D., \& {Lattanzi}, M.~G. 1997, \apjl, 476, L67

\bibitem[{{Dopita} {et~al.}(2002){Dopita}, {Groves}, {Sutherland}, {Binette},
  \& {Cecil}}]{Dopita_2002}
{Dopita}, M.~A., {Groves}, B.~A., {Sutherland}, R.~S., {Binette}, L., \&
  {Cecil}, G. 2002, \apj, 572, 753

\bibitem[{{Doucet} {et~al.}(2006){Doucet}, {Pantin}, {Lagage}, \&
  {Dullemond}}]{Doucet_2006}
{Doucet}, C., {Pantin}, E., {Lagage}, P.~O., \& {Dullemond}, C.~P. 2006, \aap,
  460, 117

\bibitem[{{Elitzur} \& {Shlosman}(2006)}]{Elitzur_2006}
{Elitzur}, M. \& {Shlosman}, I. 2006, \apjl, 648, L101

\bibitem[{{Galliano} {et~al.}(2003){Galliano}, {Alloin}, {Granato}, \&
  {Villar-Mart{\'{\i}}n}}]{Galliano_2003}
{Galliano}, E., {Alloin}, D., {Granato}, G.~L., \& {Villar-Mart{\'{\i}}n}, M.
  2003, \aap, 412, 615

\bibitem[{{Galliano} {et~al.}(2005){Galliano}, {Pantin}, {Alloin}, \&
  {Lagage}}]{Galliano_2005}
{Galliano}, E., {Pantin}, E., {Alloin}, D., \& {Lagage}, P.~O. 2005, \mnras,
  363, L1

\bibitem[{{Gallimore} {et~al.}(2004){Gallimore}, {Baum}, \&
  {O'Dea}}]{Gallimore_2004}
{Gallimore}, J.~F., {Baum}, S.~A., \& {O'Dea}, C.~P. 2004, \apj, 613, 794

\bibitem[{{Gratadour} {et~al.}(2006){Gratadour}, {Rouan}, {Mugnier}, {Fusco},
  {Cl{\'e}net}, {Gendron}, \& {Lacombe}}]{Gratadour_2006}
{Gratadour}, D., {Rouan}, D., {Mugnier}, L.~M., {et~al.} 2006, \aap, 446, 813

\bibitem[{{Greenhill} \& {Gwinn}(1997)}]{Greenhill_1997}
{Greenhill}, L.~J. \& {Gwinn}, C.~R. 1997, \apss, 248, 261

\bibitem[{{Groves} {et~al.}(2004){Groves}, {Cecil}, {Ferruit}, \&
  {Dopita}}]{Groves_2004c}
{Groves}, B.~A., {Cecil}, G., {Ferruit}, P., \& {Dopita}, M.~A. 2004, \apj,
  611, 786

\bibitem[{{H{\"o}nig} {et~al.}(2006){H{\"o}nig}, {Beckert}, {Ohnaka}, \&
  {Weigelt}}]{Honig_2006}
{H{\"o}nig}, S.~F., {Beckert}, T., {Ohnaka}, K., \& {Weigelt}, G. 2006, \aap,
  452, 459

\bibitem[{{Jaffe} {et~al.}(2006){Jaffe}, {Meisenheimer}, {Raban}, {Tristram},
  \& {R{\"o}ttgering}}]{Jaffe_2006}
{Jaffe}, W., {Meisenheimer}, K., {Raban}, D., {Tristram}, K., \&
  {R{\"o}ttgering}, H.~J.~A. 2006, in The Central Engine of Active Galactic
  Nuclei, ed. L.~{Ho} \& J.-M. S. F.~A. {Wang}

\bibitem[{{Jaffe} {et~al.}(2004){Jaffe}, {Meisenheimer}, {R{\"o}ttgering},
  {Leinert}, {Richichi}, {Chesneau}, {Fraix-Burnet}, {Glazenborg-Kluttig},
  {Granato}, {Graser}, {Heijligers}, {K{\"o}hler}, {Malbet}, {Miley},
  {Paresce}, {Pel}, {Perrin}, {Przygodda}, {Schoeller}, {Sol}, {Waters},
  {Weigelt}, {Woillez}, \& {de Zeeuw}}]{Jaffe_2004}
{Jaffe}, W., {Meisenheimer}, K., {R{\"o}ttgering}, H.~J.~A., {et~al.} 2004,
  \nat, 429, 47

\bibitem[{{Kartje} {et~al.}(1999){Kartje}, {K{\"o}nigl}, \&
  {Elitzur}}]{Kartje_1999}
{Kartje}, J.~F., {K{\"o}nigl}, A., \& {Elitzur}, M. 1999, \apj, 513, 180

\bibitem[{{Labeyrie}(1970)}]{Labeyrie_1970}
{Labeyrie}, A. 1970, \aap, 6, 85

\bibitem[{{Mason} {et~al.}(2006){Mason}, {Geballe}, {Packham}, {Levenson},
  {Elitzur}, {Fisher}, \& {Perlman}}]{Mason_2006}
{Mason}, R.~E., {Geballe}, T.~R., {Packham}, C., {et~al.} 2006, \apj, 640, 612

\bibitem[{{Nenkova} {et~al.}(2002){Nenkova}, {Ivezi{\'c}}, \&
  {Elitzur}}]{Nenkova_2002}
{Nenkova}, M., {Ivezi{\'c}}, {\v Z}., \& {Elitzur}, M. 2002, \apjl, 570, L9

\bibitem[{{Poncelet} {et~al.}(2006){Poncelet}, {Perrin}, \&
  {Sol}}]{Poncelet_2006}
{Poncelet}, A., {Perrin}, G., \& {Sol}, H. 2006, \aap, 450, 483

\bibitem[{{Rhee} \& {Larkin}(2006)}]{Rhee_2006}
{Rhee}, J.~H. \& {Larkin}, J.~E. 2006, \apj, 640, 625

\bibitem[{{Schartmann} {et~al.}(2005){Schartmann}, {Meisenheimer}, {Camenzind},
  {Wolf}, \& {Henning}}]{Schartmann_2005}
{Schartmann}, M., {Meisenheimer}, K., {Camenzind}, M., {Wolf}, S., \&
  {Henning}, T. 2005, \aap, 437, 861

\bibitem[{{Weigelt} {et~al.}(2004){Weigelt}, {Wittkowski}, {Balega}, {Beckert},
  {Duschl}, {Hofmann}, {Men'shchikov}, \& {Schertl}}]{Weigelt_2004}
{Weigelt}, G., {Wittkowski}, M., {Balega}, Y.~Y., {et~al.} 2004, \aap, 425, 77

\bibitem[{{Wittkowski} {et~al.}(2004){Wittkowski}, {Kervella}, {Arsenault},
  {Paresce}, {Beckert}, \& {Weigelt}}]{Wittkowski_2004}
{Wittkowski}, M., {Kervella}, P., {Arsenault}, R., {et~al.} 2004, \aap, 418,
  L39

\end{thebibliography}

\end{document}